\documentclass[aps, twoside, dvips, nofootinbib]{revtex4}
\usepackage{amsmath}
\usepackage{amssymb,epsfig}
\usepackage{graphics}
\usepackage{psfrag}
\usepackage{array}
\usepackage{bbm}
\newcommand{\bea}{\begin{eqnarray}}	
\newcommand{\eea}{\end{eqnarray}}
\newcommand{\be}{\begin{equation}}	
\newcommand{\ee}{\end{equation}}
\newcommand{\id}{\mathbb{I}}
\newtheorem{lemma}{Lemma}[section]
\newcommand{\tr}{\text{Tr}}

\newcommand{\rd}{\mathrm{d}}
\newcommand{\cG}{{\cal G}}
\newcommand{\bcG}{\bar{\cal G}}

\newtheorem{theorem}{Theorem}
\newtheorem{definition}{Definition}
\usepackage{subfigure}
\usepackage{color}
\begin{document}

\title{\large\bf Group field theory renormalization - the 3d case: power counting of divergences}

\author{Laurent Freidel}\email{lfreidel@perimterinsitute.ca}
\affiliation{Perimeter Institute for Theoretical Physics \\ 31 Caroline
St, Waterloo, Ontario N2L 2Y5, Canada}
\author{Razvan Gurau}\email{rgurau@perimeterinstitute.ca}
\affiliation{Perimeter Institute for Theoretical Physics \\ 31 Caroline
St, Waterloo, Ontario N2L 2Y5, Canada}
\author{Daniele Oriti}\email{daniele.oriti@aei.mpg.de}
\affiliation{Perimeter Institute for Theoretical Physics \\ 31 Caroline
St, Waterloo, Ontario N2L 2Y5, Canada \\}
\affiliation{Albert Einstein Institute, Am Muehlenberg 4, Golm,
Germany, EU}

\date{\small \today}

\begin{abstract}
We take the first steps in a systematic study of Group Field Theory renormalization, focusing on the Boulatov model for 3D quantum gravity.
We define an algorithm for constructing the 2D triangulations that characterize the boundary of the 3D bubbles, where divergences are located, of an arbitrary 3D GFT Feynman diagram.
We then identify a special class of graphs for which a complete contraction procedure is possible, and prove, for these, a complete power counting. 
These results represent important progress towards understanding the origin of the continuum and manifold-like appearance of quantum spacetime at low energies, and of its topology, in a GFT framework.\end{abstract}

\maketitle

\section{Introduction}

The field of non-perturbative quantum gravity is progressing fast, in several directions \cite{libro}. Spin foam models \cite{SF}
are one of them, and can be understood as a covariant formulation
of the dynamics of loop quantum gravity \cite{LQG} and as a new
algebraic implementation of discrete quantum gravity approaches, such as Regge calculus \cite{williams}
and dynamical triangulations \cite{DT}. The basic idea is to encode quantum gravity kinematics
in discrete quantum histories given by {\it spin foams}:
combinatorial 2-complexes labelled by group-theoretic data. The
2-complex is combinatorially dual to a simplicial complex, of which the
algebraic data determine a possible geometry. A quantum dynamics is specified by the assignment of a
probability amplitude to each spin foam, and the model is
defined by a sum over both 2-complexes and corresponding algebraic data. At present the most complete definition of a spin
foam model is achieved by means of the so-called group field
theory formalism \cite{iogft,iogft2,laurentgft}. 

\medskip

Group field
theories are quantum field theories over group manifolds, characterized by a non-local pairing of field
arguments in the action, which can be seen as a generalization of
matrix models \cite{mm} (and of
the subsequent, but less developed, tensor models
\cite{gross,ambjorn}). The combinatorics of the field arguments in
the interaction term of the GFT action follows that of (D-2) faces of a D-simplex, with the GFT field itself
interpreted as a (second) quantization of a (D-1)-simplex. The
kinetic term of the action governs the  gluing of
two D-simplices along a common (D-1)-simplex. See \cite{iogft,iogft2} for details. Because of this
combinatorial structure, the GFT Feynman diagrams, themselves cellular complexes, are dual to
D-dimensional simplicial complexes, as we will discuss at length in the following. Thus GFTs can be seen \cite{laurentgft, iogft} as a simplicial \lq\lq third quantization\rq\rq of gravity \cite{3rd}, in which a discrete spacetime emerges as a Feynman diagram of the theory in perturbative expansion. The field arguments assign group-theoretic data to these cellular complexes, and the GFT perturbative expansion in Feynman
amplitudes defines uniquely and completely \cite{mikecarlo} a spin
foam model. This makes GFTs a very useful tool, but
also suggests that they may provide a more
fundamental definition of a dynamical theory of spin networks, representing the best way to
investigate non-perturbative and collective properties of their
quantum dynamics \cite{laurentgft, iogft, gftfluid}.
The results we present in this paper are a first step in realizing this suggestion, in the simpler 3D context.

The main open problem that GFTs, as well as other discrete quantum gravity approaches, face is that of bridging the gap between their discrete description of spacetime and the one we are accustomed to at low energy, based on continuum manifolds whose geometry is governed by a classical field theory like General Relativity. As it is immediately clear, several issues are intertwined here. First of all, there is the issue of obtaining a {\it continuum} description of spacetime from the discrete structures that GFTs generate in its stead, i.e. the GFT Feynman diagrams. This means understanding in which regime of the fundamental GFT model a continuum approximation is allowed and useful to study spacetime physics. Two more technical issues are related to the fact that GFTs define a sum over simplicial complexes, representing, as we said, discrete spacetime structures, 1) of arbitrary topology and 2) not necessarily corresponding to manifolds \cite{DP-P}. In general, in fact, they fail to satisfy manifold conditions and correspond to pseudo-manifolds instead, i.e. contain conical singularities at the vertices. So one can ask why at low energy and in the continuum approximation does spacetime have a fixed (and trivial) topology and manifold properties? Both questions can and should be addressed in the GFT formalism. The first has an analogue in the context of matrix models, where it is known that diagrams of trivial topology ($S^2$ in the compact case) dominates the Feynman amplitudes of the theory in the so-called double-scaling limit \cite{mm}. The second arises only in dimensions $D>2$ \cite{DP-P}  and is known not to be easily solvable in the context of tensor models (the immediate generalization of matrix models)  \cite{gross,ambjorn}, which are characterized by trivial Feynman amplitudes, i.e. amplitudes which depend only on the combinatorics of the underlying simplicial complex. We touch on the second of these two issues, and provide some clues towards its solution, in this paper. Obviously, all these open issues are dynamical in nature, i.e. depend heavily on the quantum amplitudes of the specific GFT model one considers. This is even more true for the most important of the open problems of this approach: to reproduce a continuum manifold of some topology representing spacetime is not enough, as we want the dynamics of its geometry to be governed by General Relativity (possibly up to quantum corrections to the same). We do not touch on this last issue in this paper, and we refer to the literature for discussions, ideas and partial results concerning it \cite{laurentgft, LQG, gftfluid}.

GFTs offer a very convenient framework for investigating all the above issues, common to many different quantum gravity approaches, for a variety of reasons. An important one is that GFTs are almost ordinary quantum field theories, their only (if crucial) peculiarity being a complicated (non-local) combinatorial structure of arguments in the action and the corresponding simplicial nature of associated Feynman diagrams. Still, they allow the application of standard and very powerful quantum field theory ideas and methods to tackle quantum gravity problems, like the ones mentioned above. These methods will have to be suitably generalized to the new context, of course, and their consequences of their application re-interpreted in quantum gravity terms. 

\medskip

Among the quantum field theory methods that seem most suited to tackle these more technical issues, as well as to study the general problem of the (collective) dynamics of GFT models in different regimes, is the renormalization group.
In this paper we start a systematic study of GFT renormalization, focusing on a simple and well-known model, the Boulatov model for 3D quantum gravity \cite{boulatov}. It will allow us to develop some tools that can be later applied to other models, and to obtain a first understanding of some of the difficulties involved in applying renormalization ideas to GFTs. We will see, in fact, that this model is highly non-trivial already, and that the complicated combinatorial and topological structure of its Feynman diagrams makes each step of the usual renormalization procedures much less straightforward, but also more interesting, than in usual quantum field theories. More precisely, the divergences of this model are related to the topology of the bubbles (3-dimensional cells) in the Feynman diagrams, and a general power counting theorem is very difficult to establish mainly due to the very complicated topological structure of $3D$ simplicial complexes.

Focusing on the Boulatov model allows us also to make the first steps in understanding the role of topological invariance in GFT and in quantum gravity in general, from the point of view of renormalization. In the GFT context, that fact that the Boulatov model corresponds to a quantization of BF theory, a topological quantum field theory, translates into the property that its Feynman amplitudes depend only on the topology of the corresponding Feynman diagram, but not on its specific combinatorial structure for given topology. From the spin foam point of view, this is the well-known triangulation independence property of the Ponzano-Regge model. We expect the renormalization group to provide a new field-theoretic interpretation of this feature. more precisely, we conjecture the following scenario: topological invariance should give rise to a non-trivial fixed point of the renormalization group, and with the property that the model is dominated, in its vicinity, by simple manifold configurations. In fact, usually fixed points correspond to new symmetries, e.g. the Langmann-Szabo symmetry in non-identically distributed matrix models \cite{razvan}; also, looking again at matrix models \cite{mm}, one sees that the renormalization group identifies the most regular diagrams, planar diagrams, as the dominant ones; we look for a similar feature in the more general GFT setting, as we discuss in the following. We actually expect more: that also in higher dimensions, and for GFT models that are not themselves topological, i.e. for quantum gravity model obtained for example by constraining appropriately topological BF models, like most current spin foam/GFT models \cite{EPR,laurentkirill,gftsimplicial}, one will find non-trivial fixed points corresponding to a topological phase.

\medskip

A systematic development of GFT renormalization requires defining fist of all the GFT counterpart of the basic ingredients of the renormalization of quantum field theories. These are: a) a scale analysis; b) a locality principle; c) a power counting of divergences. A partial power counting of divergences for the Boulatov model is our main result, and it is obtained by addressing the issues of scale and of locality first.  

As far as scales are concerned, we use the spectral decomposition of the GFT propagator (a simple product of delta functions in this model) in order to introduce them easily, by an explicit cut-off. This induces a departure from topological invariance, i.e. it does not respect exactly the initial symmetries of the model, but allows to identify nicely the divergent contributions to the amplitudes (a similar cut-off procedure, which however maintains topological invariance would be to use quantum groups and work with the Turaev-Viro model \cite{boulatov, TV}).

A locality principle, in the context of renormalization, is needed only for the divergent graphs: the amplitude of such graphs must be combined with local counterterms in order to become finite. It is crucial for our purposes to note that 
in all known models, the local nature of the field theories translates into the existence of a contraction procedure for the Feynman diagrams. In usual quantum field theories, all connected graphs can be contracted to points exactly because of locality, and hence can be combined with counterterms. Similarly both in matrix models \cite{mm}, and non commutative quantum field theory models \cite{razvan1} the only divergences in the amplitudes come from planar graphs, and these are also the class of graphs that can be contracted, thus are in some sense local, so that once more they can be combined with counterterms and renormalized. This suggests that the point of view on locality, diagram contraction and renormalization can be reversed. Take for instance an arbitrary matrix model, not necessarily identically distributed. Even before establishing a power counting theorem, the following question is legitimate: is there a family of graphs which can be fully contracted? The answers is that planar graphs are the only ones for which a complete and consistent contraction procedure can be defined. Hence, if a matrix model is to allow for the application of renormalization group procedure and thus be renormalized in the standard sense, all its divergent, and thus the only relevant, graphs must be planar. However, we know very well by now that, in order to reach the point where renormalization can be applied, i.e. in order to be in the regime in which only planar graphs are relevant, for a given matrix model, one might need further ingredients. For instance, for a model of identically distributed matrices one must introduce a scaling limit \cite{mm}.
We apply this reasoning to the 3D Boulatov GFT model, and ask: is there a family of graphs on which one could define a  full contraction procedure? In light of our proof of power counting we can give a definite answer. We show that the only family of graph for which a full contraction procedure exists are the ``type 1'' manifolds defined in section \ref{sec:bubbles}. Hence, in order to be able to define a renormalization group for this model, all divergent diagrams better be ``type 1''. The need for a contraction procedure in the context of GFT renormalization also explains why we do not try to define a simpler, global proof for the degree of divergence of GFT diagrams, and look instead for one based on this contraction, even if, as we will see, it turns out to be rather involved.

\medskip

Anyway, as anticipated, for these \lq\lq type 1\rq\rq diagrams, we are able to define this complete contraction procedure and prove a corresponding power counting theorem for their divergences. As a first step in the process of defining the contraction and identifying divergences, we will give an algorithm for constructing the 2D triangulations that characterize the boundary of the 3D bubbles of an arbitrary 3D GFT diagram graph. We believe that this algorithm is an interesting and useful result in itself. As mentioned, it is in fact at these bubbles that the divergences of the model are located. However, as we will see, and as it is the case with matrix models, the power counting we establish is not uniform in the number of internal vertices. We then conjecture that in order to define a renormalization group for GFT models we again need to take an appropriate scaling limit. We also note that this result can not be extended to arbitrary manifold configuration (as it will be shown on a counterexample).

\section{The Boulatov model revised}

We start defining the model, i.e. the group field theory defined by Boulatov \cite{boulatov} for 3d Riemannian quantum gravity. Like in any other group field theory, each field, here a function $\phi(g_1,g_2,g_3): SU(2)^{\times 3} \rightarrow \mathbb{R}$, is interpreted as representing a (D-1)-simplex, here a triangle, and the crucial feature of the action is that the combinatorics of field arguments in the kinetic and interaction term is chosen in such a way that the Feynman diagrams of the theory are cellular complexes that are topologically dual to simplicial complexes, 3-dimensional ones in this specific case. The Feynman amplitudes of this field theory, i.e. the corresponding spin foam model, are products of delta functions, each of them  associated to a face of the Feynman diagram representing a discrete spacetime. The arguments of these a delta function are the holonomies of an $SU(2)$ connection associated to the face. The model reproduces, in its perturbative expansion, the (trivial) geometric content of 3d gravity discretized on the simplicial complex dual to each Feynman diagram (flatness of the gravitational connection). The corresponding spin foam formulation is the well-known Ponzano-Regge model \cite{PR1}.

In order to simplify our analysis we will look at an ``orientable'' version of the Boulatov model (meaning that the simplicial complexes dual to its graphs are orientable). 

\begin{figure}[htb]
\centering
\subfigure[Labeled tetrahedron.]{\label{fig:tetrahedron}
\includegraphics[width=40mm]{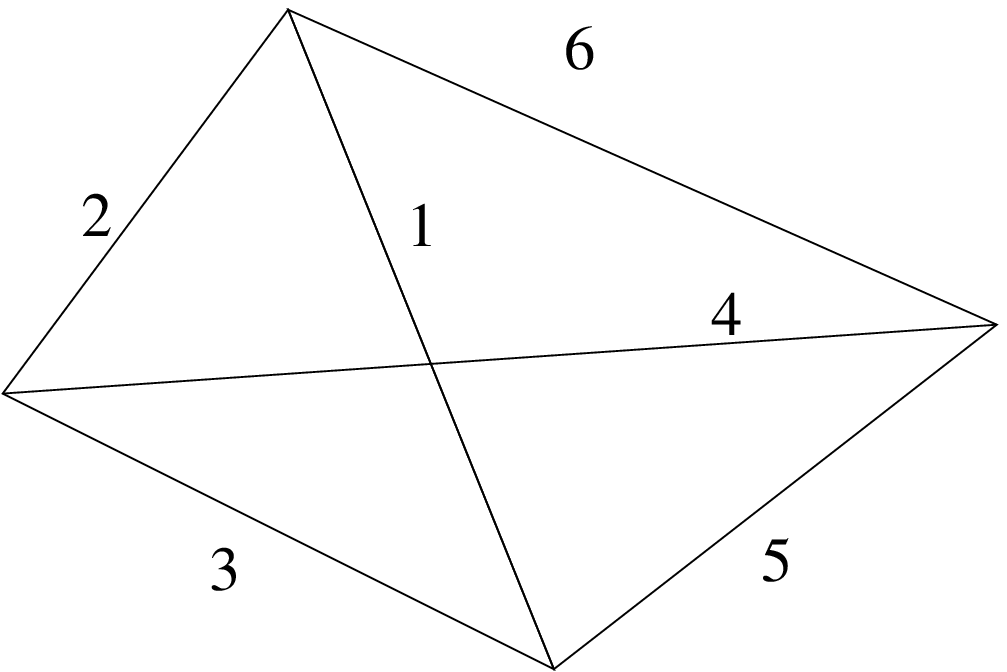}}
\hspace{3cm}
\subfigure[Labeled four valent vertex.]{ \label{fig:vertexsimpl}
\includegraphics[width=40mm]{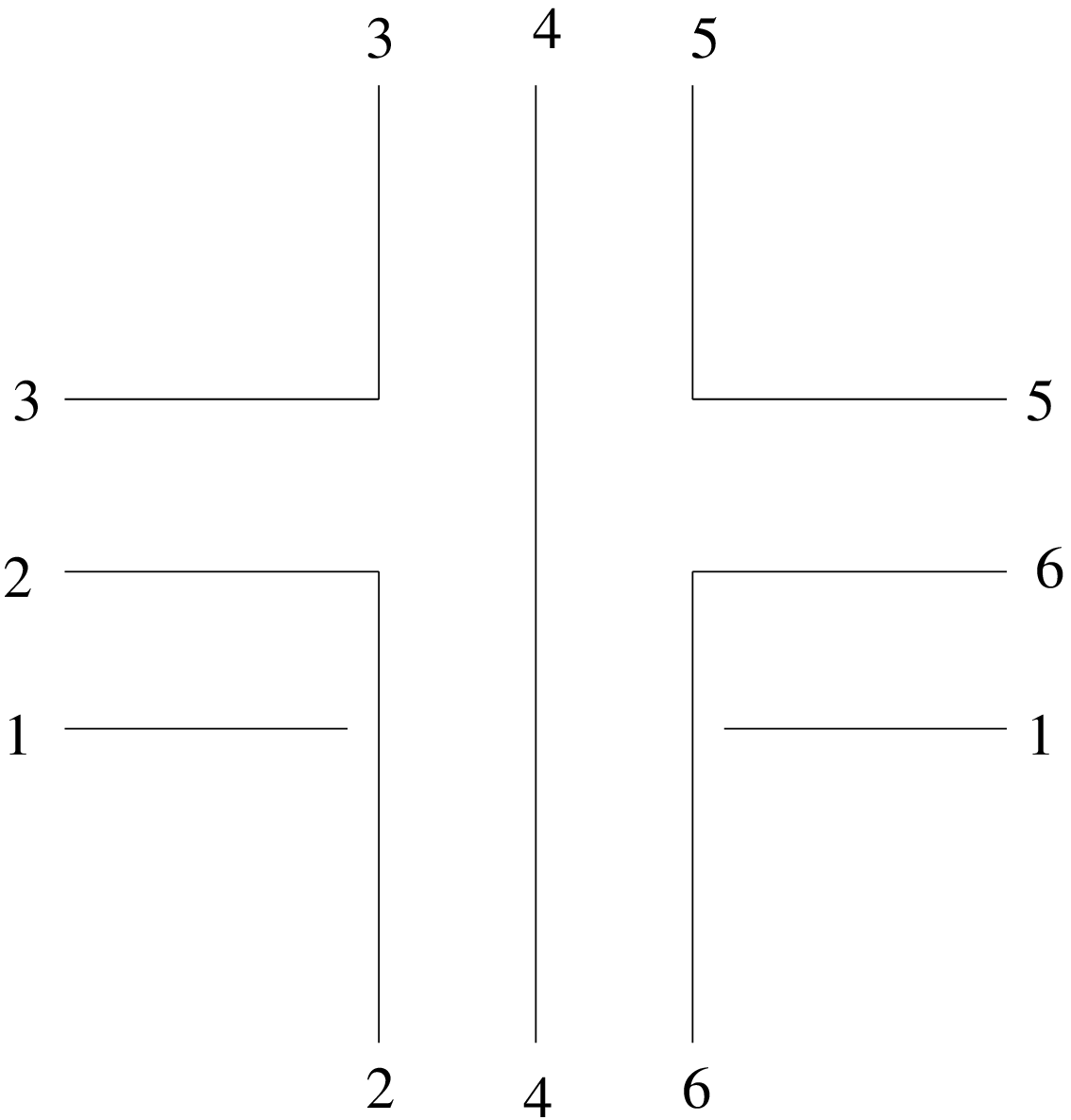}}
\caption{The GFT vertex.}
\end{figure}

We consider a tetrahedron labeled as in figure 
\ref{fig:tetrahedron}, and we orient all triangles consistently with the exterior normals of the tetrahedron. Thus the four oriented triangles are $(1,2,3)$, $(3,4,5)$, $(5,6,1)$ and $(6,4,2)$, and these become the labeling of arguments of the four fields in the interaction term of the action.  Notice that taking into account the orientation
of the tetrahedra results in a specific pairing of the group variables (field arguments) 
in the interaction term. Moreover this pairing is {\it different} from the naive pairing which is usually 
prescribed. As said, our choice is dictated by the fact that we want all normals associated to the triangles of the tetrahedron oriented outwards. The dual of the tetrahedron is the 4-valent vertex of the GFT action. Labeling the vertex clockwise we end up with the picture drawn in figure \ref{fig:vertexsimpl}. The lines in the vertex ($1,2,\dots 6$) will henceforth be referred to as strands.

\begin{figure}[htb]
\centering
\subfigure[Labeled triangles.]{\label{fig:triang}
\includegraphics[width=40mm]{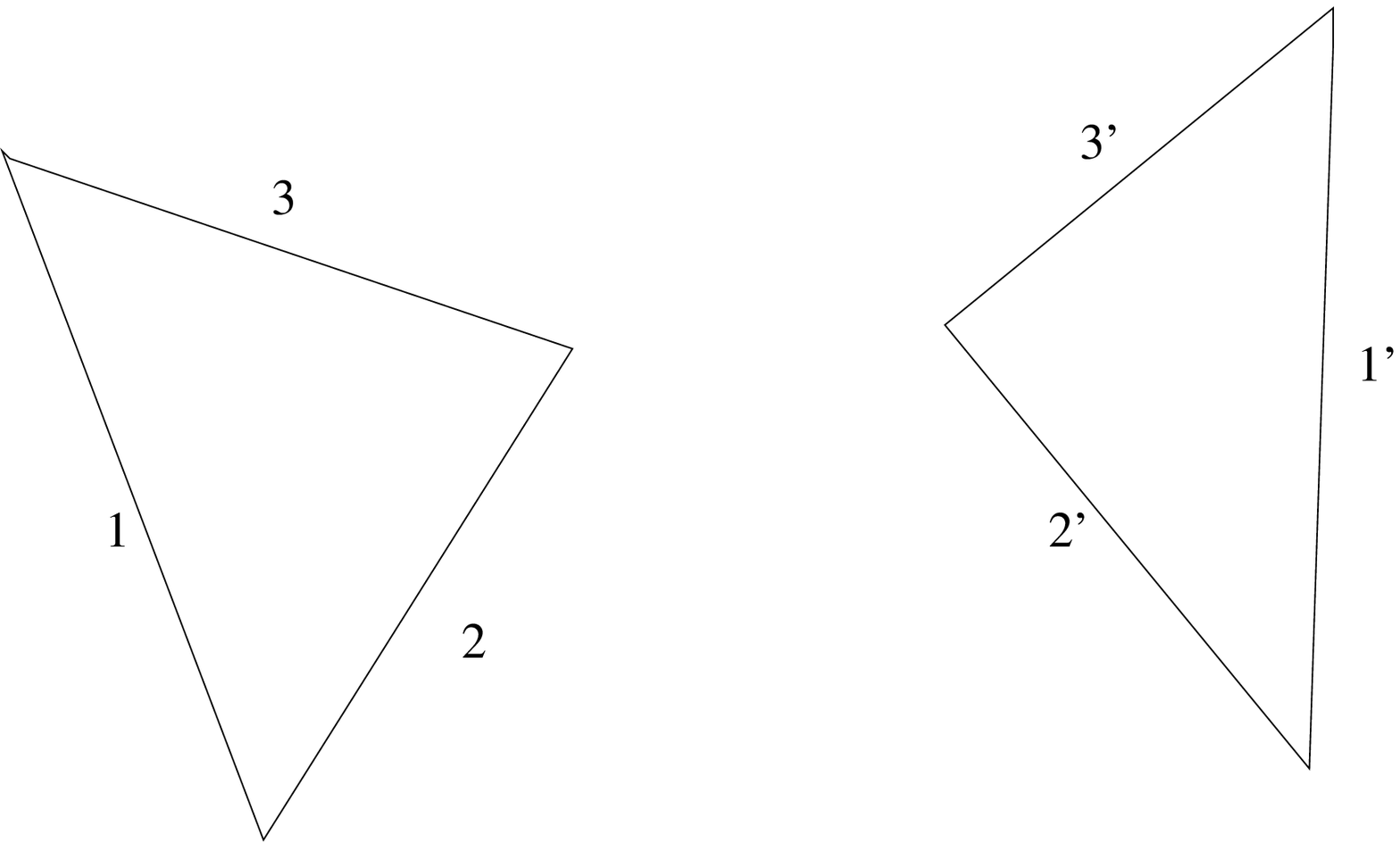}}
\hspace{3cm}
\subfigure[Labeled propagator.]{ \label{fig:propsimpl}
\includegraphics[width=40mm]{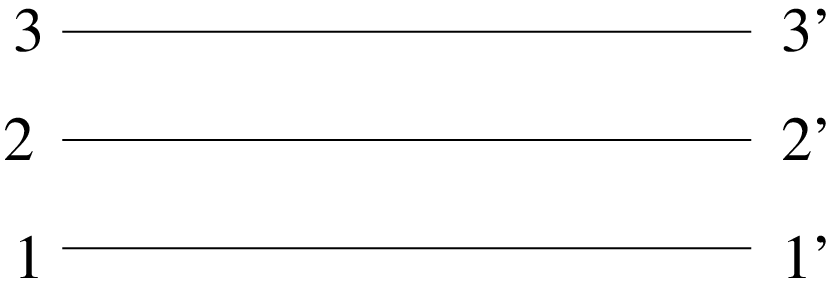}}
\caption{The GFT propagator.}
\end{figure}

Two tetrahedra are glued along triangles. The identification of the two triangles $(1,2,3)$ and $(3',2',1')$ in figure \ref{fig:triang} gives $1=1'$, $2=2'$ and $3=3'$. This identification is reproduced by the GFT propagator, represented in \ref{fig:propsimpl}.

\medskip

Thus the GFT whose graphs are orientable in the sense above is given by the action
\bea
S\,=\,\frac{1}{2} \int dg_1dg_2dg_3 \phi(g_1,g_2,g_3)\phi(g_3,g_2,g_1) 
\,+\,\frac{\lambda}{4} \int \phi(g_1,g_2,g_3)\phi(g_3,g_4,g_5) \phi(g_5,g_6,g_1)
\phi(g_6,g_4,g_2) \; ,
\eea
where the field $\phi$, that we take as real-valued, is {\bf not} assumed to have specific symmetry properties under permutations of its three arguments; that is, unlike \cite{DP-P,DFKR} we associate to each line in a Feynman diagram only the identity permutation. The integrations over the group (left implicit in the interaction term) are performed with the invariant Haar measure.

The propagator of our model is thus 
\bea
P(g_1,g_2,g_3;g_{3'},g_{2'},g_{1'})=\int dh \; \delta(g_1 h g_{1'}^{-1}) \delta(g_2 h g_{2'}^{-1}) \delta(g_3 h g_{3'}^{-1})\; ,
\eea
imposing the simple identification of arguments we mentioned.

The model is then defined, at the quantum level, by the partition function, expanded in perturbation theory as:

$$ Z\,=\,\int
\mathcal{D}\phi\,e^{-S[\phi]}\,=\,\sum_{{\cal G}}\,\frac{\lambda^N}{\mathrm{sym}[{\cal G}]}\,A({\cal G}),
$$
where $N$ is the number of interaction vertices in the Feynman
graph ${\cal G}$, $\mathrm{sym}[{\cal G}]$ is a symmetry factor for the graph
and $A({\cal G})$ the corresponding Feynman amplitude. 

Having identified the propagator and vertex function of the model, we can now construct the Feynman amplitudes. However, these being given by products of delta functions on the $SU(2)$ group, some regularization is needed in order to have them well-defined.  There exists several regularizations of the Boulatov model. The best known is the Turav-Viro model, obtained by switching from $SU(2)$ to its quantum deformation $SU(2)_q$ with $q$ a root of unity \cite{boulatov,TV}; this model is known to be related to 3d riemannian quantum gravity with positive cosmological constant; the quantum deformation has the immediate effect of restricting the representation summed over in the mode expansion of the Feynman amplitudes to a finite range, thus imposing an infra-red regulator (associated to the cosmological constant). Another possible regularization is obtained by substituting each delta function in the propagator with a heat kernel on the group, with (temperature) parameter $\beta$; this corresponds to leaving the range of representations summed over in momentum space unrestricted, but inserting in it a regularizing factor $e^{\frac{1}{\beta}j(j+1)}$; for $\beta\rightarrow\infty$ one recovers the original model. Looking instead directly at the Feynman (spin foam) amplitudes of the model, one can use another regularization that involves inserting appropriate factors associated to vertices to the simplicial complex dual to the Feynman diagram \cite{laurentdiffeo,PR1,LCS, barrett}. These factors can be interpreted \cite{laurentdiffeo,LCS} as the volume of the gauge group associated to the translation symmetry of discretized BF, and this regularization becomes then a gauge fixing procedure for this symmetry. It is not clear, however, how to implement this last regularization procedure at the GFT level, ultimately because we do not know yet how to identify the translation symmetry at the level of the Boulatov action.  Here we choose a different one. Let $\Lambda$ be a large number, we define the cutoffed propagator like
\bea
P^{\Lambda}(g_1,g_2,g_3;g_{3'},g_{2'},g_{1'})=\int dh \; \delta^{\Lambda}(g_1 h g_{1'}^{-1}) \delta^{\Lambda}(g_2 h g_{2'}^{-1}) \delta^{\Lambda}(g_3 h g_{3'}^{-1})\; ,
\eea
with
\bea
\delta^{\Lambda} (h)=\sum_{j=0}^{\Lambda} d_j\chi^j(g)\, .
\eea
The nice feature of this regularization is that, on top of being very simple, it preserves the composition properties of the delta functions (as the heat kernel regularization does), and also allows easy evaluations of amplitudes (which are instead a bit more complicated with the heat kernel regularization).   
Using the explicit form of our regularized $\delta$ functions we can prove, for example, the following two properties\footnote{which follow from $\int dh D^{j'}_{m'n'}(h) \bar {D}^{j}_{mn}(h)=\frac{1}{d_j}\delta^{jj'}\delta_{m'm} \delta_{n'n}.$}
\bea
\int dh\delta^{\Lambda}(gh^{-1})\delta^{\Lambda}(hg'^{-1}) 
&=&\delta^{\Lambda}(gg'^{-1})\,,\\
\int dh \delta^{\Lambda}(gh)&=& 1 \; .
\eea

A graph ${\cal G}$ of this model is formed of the vertices and propagators drawn in 
\ref{fig:vertexsimpl} and \ref{fig:propsimpl}. We denote ${V}_{\cal G}$ the set of internal vertices (label $V_1,\dots V_{|{V}_{\cal G}|}$, ${ L}_{\cal G}$ 
the set of internal lines (labeled $L_1,\dots L_{|{ L}_{\cal G}|}$). The closed circuits in the graph correspond to faces. We denote ${ F}_{\cal G}$ the set of internal faces (labeled $F_1,\dots F_{|{\cal F}_{\cal G}|}$). Finally, closed three dimensional regions of the graph are called bubbles. We denote ${ B}_{\cal G}$ the set of internal bubbles (labeled $B_1,\dots B_{|{ B}_{\cal G}|}$).

In order to write the amplitude of a graph we chose an orientation for each of its lines and faces. The amplitude of a graph then writes (in self explaining notations) as
\be\label{AG}
\lambda^{| {V}_{\cal G}|} A({\cal G})=\lambda^{|{V}_{\cal G}|}\left( \int \prod_{L\in { L}_{\cal G}} dh_L \prod_{F\in { F}_{\cal G}} \delta^{\Lambda}\left({\overleftarrow \prod}_{L \in \partial F} h_L\right)  \right)
\ee
with $h_L$ or $h_L^{-1}$ chosen in the argument of the $\delta$ function corresponding to the face $F$ according to whether the orientation of $F$ coincides or not with that of the line $L$.
The arrow over the product express that the product is an order product over the lines in $\partial F$ in the order that 
is induced by the orientation of $F$.

\section{Bubbles}\label{sec:bubbles}
\subsection{Labeling the bubbles}
Each Feynman diagram will be given by construction, as we said, by a cellular complex dual to a 3-dimensional simplicial complex. 
The vertices, lines and faces of this graph ${\cal G}$, dual respectively to tetrahedra, triangles and edges of the simplicial complex, are readily identified. On the contrary the 3-cells, also called \lq\lq bubbles\rq\rq and dual to vertices of the simplicial complex, are difficult to identify. In this section we present the algorithm which allows one to identify them, first of all, and construct a graph which characterizes the combinatorial structure of each bubble of a generic graph ${\cal G}$. This will be crucial for our contraction procedure, and for establishing our power counting theorem. It also allows to determine easily the topology of the boundary of each bubble of ${\cal G}$, and let us stress that identifying this boundary topology is the ingredient needed to determine whether the simplicial complex dual to each graph ${\cal G}$ is a simplicial manifold or not. Indeed, the necessary and sufficient condition for such a simplicial complex to be a manifold is that every one of its bubbles has the topology of a 3-ball, i.e. that their boundaries are 2-spheres.

\begin{figure}[htb]
\centerline{
\includegraphics[width=60mm]{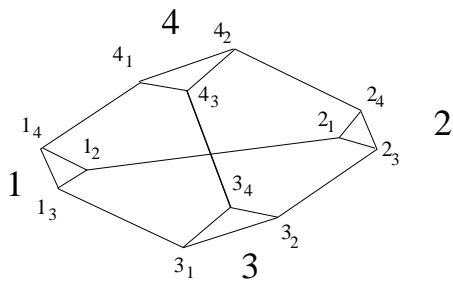}}
\caption{Fully labeled truncated tetrahedron}\label{fig:tetracarv}
\end{figure}

The basic idea is the following. Since bubbles of ${\cal G}$ are dual to the vertices of its dual simplicial complex, we can define a triangulation of their boundary in such a way that each tetrahedron in the simplicial complex corresponds to a triangle.
The way to visualize it is that we carve out a small spherical neighborhood of each vertex of the initial triangulation.
This corresponds to carving  out each corner of the tetrahedra  like in figure \ref{fig:tetracarv}. 
Gluing back these truncated tetrahedra we obtained a 3d triangulation with a 2d boundary which is also triangulated.

The topological elements of the tetrahedron, and of the simplicial complex it belongs to, are referred to as ``3 dimensional''. They are labeled as follows (see figure \ref{fig:tetracarv})
\begin{itemize}
\item The vertices of the tetrahedron (3D vertices) are labeled by a number, $1$, $2$, $3$ and $4$.

\item The edges of the tetrahedron (3D edges) connect two vertices, hence will be denoted by un-ordered couples of numbers\footnote{ the edge orientation is irrelevant because each 3D edge is assigned a different orientation in the two triangle it belongs to.}. For instance, the line connecting the vertices $1$ and $2$ is denoted $(1,2)$, etc...

\item The triangles of the tetrahedron (3D triangles) are formed by three vertices, and three edges. We denote them by ordered triples of numbers (associated to vertices) $(3,4,1)$, $(3,1,2)$, $(3,2,4)$ and $(4,2,1)$. Each triangle has  boundary identified by a set of edges. For instance, the edges in the boundary of the triangle $(3,4,1)$ are $(3,4)$, $(4,1)$ and $(1,3)$ .

\item The tetrahedron itself (3D tetrahedron) is denoted $(1,2,3,4)$, i.e. it is identified by the unordered set of its four vertices.
\end{itemize}

In order to identify the bubbles we must then deal with two distinct simplicial complexes. In fact, besides the the 3D simplicial complex of tetrahedra glued along their boundary triangles, we need to consider also a second simplicial complex, made of the small triangles carved in figure \ref{fig:tetracarv} glued along their boundary edges.
We now explain a convenient labeling of the elements of this second simplicial complex and their relation with the original 3D one. 
The topological elements of the small triangles are called ``2 dimensional'' (2D) and are labeled as follows (see figure \ref{fig:tetracarv})
\begin{itemize}
 \item The vertices of the small triangles (2D vertices) are labeled by a number and an index. 
The number keeps track of the 3D vertex of the tetrahedron to which the small triangle is associated, 
while the index indicates the 3D edge of the tetrahedron transverse to the 2D vertex. Thus the 2D vertex 
 $1_2$ belongs to the triangle corresponding to the 3D vertex $1$, and transverse to the 3D edge 
$(1,2)$, while the 2D vertex $1_3$ belongs to the triangle associated with the 3D vertex $1$ and is touched by the 3D edge $(1,3)$.
\item The edges of the small triangles (2D lines) are labeled by a couple of 2D vertices. Thus $(1_2,1_3)$ is the edge going from $1_2$ to $1_3$.

\item The small triangles (2D triangles) are labeled by there vertices $(1_2,1_3,1_4)$, $(2_1,2_4,2_3)$, $(3_1,3_2,3_4)$ and $(4_1,4_3,4_2)$; these 2D triangles are oriented, again with outward normals.

\end{itemize}

All the topological elements of the 2D complex (connected components, vertices, edges, triangles) are obtained by ``projecting'' those of the 3D simplicial complex (vertices, edges, triangles, tetrahedra)

Consider the projection of the elements of a single tetrahedron. From figure \ref{fig:tetracarv} we see that
\begin{itemize} 
\item The oriented 3D tetrahedron $(1,2,3,4)$ projects into  {\bf four} oriented 2D triangles $(1_2,1_3,1_4)$, $(2_1,2_4,2_3)$, $(3_1,3_2,3_4)$ and $(4_1,4_3,4_2)$.

\item Each 3D triangles (i-e $(3,4,1)$,  $(3,1,2)$, $(3,2,4)$ or $(4,2,1)$ which are faces of the 3D tetrahedron)  project into {\bf three} oriented 2D edges. Thus the triangle $(3,4,1)$ projects into the lines $(3_4,3_1)$, $(4_1,4_3)$ and $(1_3,1_4)$, while $(3,1,2)$ projects into $(3_1,3_2)$, $(1_2,1_3)$ and $(2_3,2_1)$, and so on.

\item Each one of the six 3D edges of the tetrahedron $(1,2)$, $(1,3)$, $(1,4)$, $(2,3)$, $(2,4)$, and $(3,4)$ project each into 
{\bf two} 2D vertices. $(1,2)$ projects into the vertices $1_2$ and $2_1$, $(1,3)$ projects into the vertices $1_3$ and $3_1$, and so on. 

\item Each vertex of the 3D tetrahedron $1$, $2$, $3$ and $4$ projects to {\bf one} 2D connected component (a 2D triangle).

\end{itemize}

The 3D topological elements are called ``ancestors'' of those obtained by projection to the 2D simplicial complex, 
(and the latter are the descendents of the former). A 3D tetrahedron has  for instance four 2D triangles descendents, 
a 3D triangle has three 2D edges descendents, and so on.

The two simplicial complexes are dual to graphs. The 3D complex is dual to a 3D GFT graph ${\cal G}$, while the 2D simplicial complex is dual to the 2D graph, called $\bar{\cal G}$.
A 3D GFT graph ${\cal G}$ is a fat graph carrying three strands per edges while a 2D graph is a fat graph carrying two strands per edges.

\begin{figure}[htb]
\centerline{
\includegraphics[width=60mm]{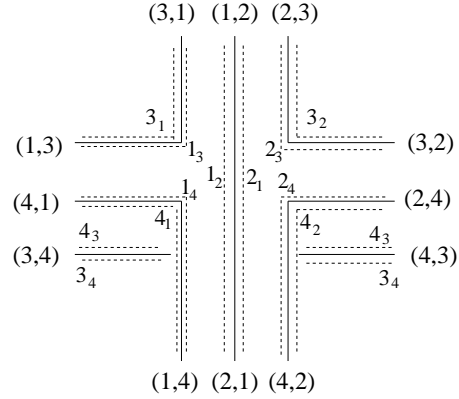}}
\caption{Fully labeled 3D interaction (dual vertex)}\label{fig:vertex}
\end{figure}

The dual of a tetrahedron is the GFT vertex. 
In the triangulation representation each  3D edges projects into two vertices in the 2D complex. 
The 3D edges of the tetrahedron are dual to faces made up of 3D strands, 
whilst the 2D vertices are dual to faces made up of 2D strands. 
Consequently in the graph picture, each 3D strand in the 3D dual vertex projects into two 2D strands, one on the left and one on the right.
The fully labeled dual of the tetrahedron from figure \ref{fig:tetracarv} is drawn in figure \ref{fig:vertex}, with the 2D strands represented by dashed lines.

We can now translate trivially the projection at the level of 3D and 2D graphs.
\begin{itemize}
 \item The 3D dual vertex $(1,2,3,4)$ projects into {\bf four} 2D dual vertices 
$(1_2,1_3,1_4)$, $(2_1,2_4,2_3)$, $(3_1,3_2,3_4)$ and $(4_1,4_3,4_2)$.
 \item Each one of the four 3D dual halflines $(3,4,1)$, $(3,1,2)$, $(3,2,4)$ and $(4,2,1)$  projects into 
{\bf three} 2D dual halflines. $(3,4,1)$ projects into the $(3_4,3_1)$, $(4_1,4_3)$, and $(1_3,1_4)$, etc.
\item Each one of the 3D strands $(3,4)$, $(4,1)$, $(1,3)$, $(1,2)$, $(2,3)$, and $(2,4)$ projects into {\bf two} 2D strands.$(3,4)$ projects into $3_4$ and $4_3$, $(4,1)$ projects into $4_1$ and $1_4$, etc. 
\item Each 3D bubble projects into {\bf one} 2D connected component (connected surface whose dual is made out of 2D triangles); this 2D connected component, as we have noted above, characterizes the topology of the boundary of the bubble, and thus allows the determine if the 3D simplicial complex satisfies manifold conditions or not.
\end{itemize}

The 3D graph ${\cal G}$ is formed of 3D vertices and 3D lines. The 3D lines are identifications of two 3D half lines. The 2D descendents of a 3D line are then obtained by identifying  the 2D descendents of the two 3D half lines.

\begin{figure}[hbt]
\centerline{
\includegraphics[width=100mm]{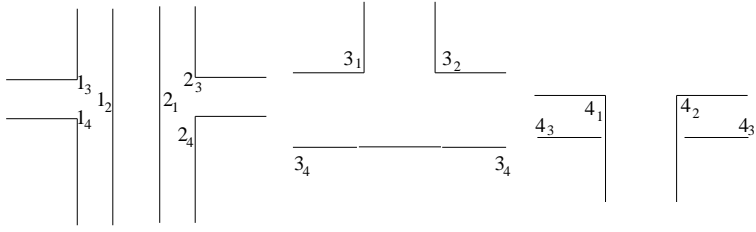}}
\caption{2D dual vertices descendent from a 3D dual vertex.}\label{fig:vertex1}
\end{figure}

\begin{figure}[hbt]
\centerline{
\includegraphics[width=100mm]{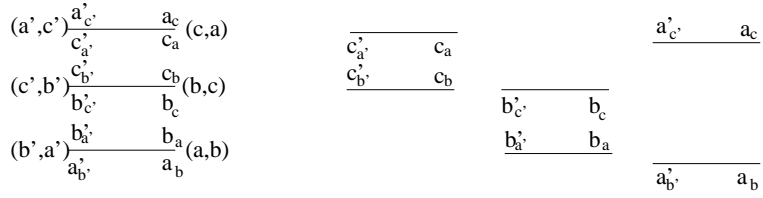}}
\caption{2D dual lines descending from a 3D dual line.}\label{fig:propag}
\end{figure}

The algorithm to draw the 2D projection $\bar {\cal G}$ of a 3D graph ${\cal G}$ is then the following.
Start by drawing the 2D descendents of all 3D dual vertices. Each 3D dual vertex will give four such descendents (see figure \ref{fig:vertex1}). Draw all descendents of the 3D dual lines. Each 3D dual line will have three descendents (see figure \ref{fig:propag}). Connect the 2D dual vertices using the 2D dual lines. The 3D dual faces of the graph ${\cal G}$ are closed 3D strands while the 2D dual faces of $\bar {\cal G}$ are  closed 2D strands.

\subsection{Characterizing the bubbles} 
From now on, we stop labeling elements of ${\cal G}$ and $\bar{{\cal G}}$ as \lq\lq dual\rq\rq for simplicity. This can be done without confusion because, from now on, 
we refer  only on the 3D and 2D graphs ${\cal G}$ and $\bar{\cal G}$,
 leaving aside their corresponding simplicial complexes.

Let
${V}_{\cal G}$, ${L}_{\cal G}$, ${F}_{\cal 
G}$, ${B}_{\cal G}$ the sets of vertices, lines, faces and bubbles of the 3D GFT graph ${\cal G}$.
Similarly we denote by 
${v}_{\cal G}$, ${l}_{\cal G}$, ${f}_{\cal 
G}$, ${b}_{\cal G}$ the sets of vertices, lines, faces and connected components, of the 2D graph  $\bar {\cal G}$ obtained by the  projection procedure.  We then have the relations:

\bea\label{eq:3D2D}
|{v}_{\cal G}|=4 |{V}_{\cal G}| \quad
|{l}_{\cal G}|=3 |{ L}_{\cal G}| \quad
|{f}_{\cal G}|=2 |{F}_{\cal G}| \quad
|{b}_{\cal G}|=|{ B}_{\cal G}| \;.
\eea

A first consequence of the above equalities is the following. For any 2D graph $\bar {\cal G}$ we can express the Euler characteristic of the 
2D surface it represent in terms of an alternating sum of elements:
\bea\label{eq:euler}
|v_{\cal G}|-|{l}_{\cal G}|+|{f}_{\cal G}|=2|{b}_{\cal G}|-
2\sum_{b\in{b}_{\cal G}}g_b \; ,
\eea
with $g_b$ the genus of the connected component $b$ (which is is the boundary of some given bubble of the 3D graph ${\cal G}$). 

Suppose that the 3D graph ${\cal G}$ is a vacuum graph.
Since the 3D GFT graph is 4-valent we have $2|{ V}_{\cal G}|=|{ L}_{\cal G}|$. Substituting eq. (\ref{eq:3D2D}) into eq. (\ref{eq:euler}) and using this relation yields
  the identity
\bea
|{V}_{\cal G}|-|{ L}_{\cal G}|+|{ F}_{\cal G}| - |{ B}_{\cal G}|=-\sum_{B\in { B}_{\cal G}}g_B \;,
\eea
with $g_B$ the genus of the boundary of the bubble $B$.
The LHS of this identity is the Euler characteristic of the simplicial complex dual to the 3D GFT graph.
It is zero if and only if this simplicial complex is a simplicial manifold.
Equivalently this means that a  3D closed graph represents a manifold 
if all the bubbles are spherical ($g_{B}=0$).
This is a well known condition that insures that the neighborhood of each vertex of the triangulation
is isomorphic to a three ball. If one of the bubble dual to a vertex is {\it not} spherical the neighborhood of this vertex contains a non contractible 
torus or genus $g$ surface and the corresponding simplicial complex is only a pseudo-manifold.

The purpose of this work is to understand better the renormalization properties of the amplitude (\ref{AG}).
As can been seen from its definition, The divergences are 
related to the presence of the delta function for each face, while naively 
each integration can potential kill one delta function.
A closer look shows that because the integrand is invariant under gauge transformations 
acting at the vertices of the graph not all the integration over $g_{L}$ can kill the delta functions only 
at most $|L_{\cal G}| - |V_{G}| +1$.
Thus it looks that the behavior of the integral 
is characterized by a naive degree of divergence given by: $
|F_{\cal_{G}}|- |L_{\cal G}| + |V_{G}| -1 = B_{\cal G } -\sum_{B\in { B}_{\cal G}}g_B.$

This is only a very rough estimate of the behavior of this integral.
In fact, it is easy to see that if one can isolate a bubble of genus $g$ from the others, then 
it carries a summation $ \sum_{j} \mathrm{ d}_{j}^{2(1-g)} =\int \prod \rd a_{i}\rd b_{i} \delta^{\Lambda}(\prod_{i=1}^{g}[a_{i},b_{i}]) $ 
Thus the spherical bubbles contributes more that the toric bubbles while an isolated genus $g>1$ bubbles gives a convergent contribution.

Our goal is now to devise a new criterion analogous to the criterion of planarity in 2D GFT models or in matrix models that allow one to identify a subclass of graphs which dominates the contribution of the path integral in the limit of large spins. 

Let's take a step back and revise the similar problem in matrix models \cite{mm}. Suppose that we start with the naive action
\bea
S=\frac{1}{2}\; \tr H^2 +\frac{\lambda}{4} \;\tr H^4 \;,
\eea
for $H$ some hermitian $\Lambda\times \Lambda$ matrix. For a given graph ${\cal G}$, denote $n,l,f$ its numbers of vertices lines and faces. Its amplitude is
\bea
A_{\cal G}=\lambda^n\Lambda^f \; .
\eea
As this amplitude is not uniform in the number of internal vertices one could naively conclude that no renormalization transformation can be defied for a matrix model.

However this is known to be false. If one wishes to obtain an uniform power counting, not depending on the number of internal vertices, one needs to consider a rescaling of the coupling constant by some power of the cutoff
\bea
\lambda\rightarrow \frac{\lambda'}{\Lambda^{\alpha}}
\eea

In order to determine the appropriate rescaling power $\alpha$, recall that there exists a family of graphs for which a contraction procedure can be defined: the planar graphs. The combinatorics of the amplitude subtraction of a renormalization transformation can only be satisfied by this family.

The lowest order planar graph reproducing the vertex is made by two vertices connected by two lines and two line which trap an internal face. It has amplitude $\lambda^2\Lambda$. 
Requiring that the graph ${\cal G}$ and all the graphs where some of its vertices have been replace by such lowest order planar insertions have the same behavior with the cutoff, one determines the appropriate power $\alpha$
\bea
\frac{\lambda'}{\Lambda^{\alpha}}=\frac{\lambda'^2}{\Lambda^{2\alpha}}\Lambda\Rightarrow \alpha=1
\eea

A posteriori one notes that after rescaling, ${\cal A}_{\cal G}=\lambda'^n \Lambda^{n-f}=\lambda'^n\Lambda^{2-2g}$, hence the planar graphs truly dominate the partition function.

The situation is more involved in 3D GFT. Still, one can start by looking for a family of graph for which a contraction procedure can be defined. Once such a family is identified, by appropriately choosing a rescaling, one can presumably construct a model where such graphs dominate. The family of contractible graphs, also called type 1, is defined below and thoroughly analyzed in the rest of our paper.

Let $b$ a 2D connected component of the 2D projected graph $\bar {\cal G}$, and let ${\cal G}^b$ the 3D graph formed by the 3D ancestors of all its vertices and lines and $\bar{\cal G}^b$ the 2D the projection of ${\cal G}^b$.
 
\begin{definition} \label{def:type1}
 A graph ${\cal G}$ is called ``type 1'' if
  \begin{itemize} 
 \item  $\forall \; b$, any 3D vertex of ${\cal G}^b$ (and consequently 3D line or 3D face) projects 
into an {\bf unique} 2D vertex (line or face) of $b$.
  \item For all $b_1$, there exists an ordering $b_1<b_2\dots < b_{|{ B}_{\cal G}|}$ such that
\begin{itemize}
\item  $b_i \cap \bar{\cal G}^{b_1\cup \dots \cup b_{i-1}}$ is connected, for all$\;i \le |{ B}_{\cal G}|$.
\item $ b_i$ is not  contained  in   $\bar{\cal G}^{b_1 \cup b_2...\cup b_{i-1}}$, for $\;i < |{ B}_{\cal G}|$..
\end{itemize}
\end{itemize}
\end{definition}

We refer to the ordering $b_1<b_2<\dots < b_{|{\cal B}_{\cal G}|}$ as the ordering associated to $b_1$.
We sometimes refer to a graph which is not ``type 1''  as a ``type 2'' graph.

In a later section we establish the result 
\begin{theorem}
 The amplitude of a connected ``type 1'' {\it manifold} vacuum 3D graph ${\cal G}$ is
   \bea
    A_{{\cal G}}\,=\,
   \lambda^{|{V}_{\cal G}|}\left( \delta^{\Lambda}(\id)\right)^{|{B}_{\cal G}|-1}
   \eea
\end{theorem}

Both conditions in definition \ref{def:type1} enter crucially in our proof. Furthermore, the result is achieved by a complete contraction procedure. In this sense, the ``type 1'' graphs are the family of contractible graphs. Any renormalization transformation must subsequently subtract only ``type 1'' graphs.

We note that again, for the naive 3D GFT model, the degree of divergence is not uniform in the number of internal vertices. Inspired by the previous discussion on matrix models, we look for some rescaling of the coupling constant which presumably would render this degree of divergence uniform.

To find the appropriate power of the rescaling, we consider the first ``type 1'' graph reproducing a vertex. It is the graph depicting a one to four topological move on the vertex. Its amplitude is $\lambda^4\delta(\id)\approx \lambda^{4}\Lambda^3$. Requiring again that the degree of divergence of a graph is invariant under substituting any vertex with this lowest order ``type 1'' insertion yields
\bea
\frac{\lambda'}{\Lambda^{\alpha}}=\frac{\lambda'^4}{\Lambda^{4\alpha}}\Lambda^3\Rightarrow \alpha=1
\eea

A posteriori we note that after rescaling the amplitude of a type 1 graph is 
$A_{{\cal G}}=\lambda'^{|{V}_{\cal G}|} \Lambda^{-|{V}_{\cal G}|+3({B}_{\cal G}|-1)}$.

For this scaling limit to exist and behave in a similar way to that of our previous toy model, the following conjectures (amply verified on examples) must hold

{\bf Conjecture 1:} At fixed order of perturbation, the number of bubbles $|{B}_{\cal G}|$ is 
maximal for type 1 graphs, and strictly smaller for all other graphs. If this holds, then the ``type 1'' graphs dominate the partition function.

{\bf Conjecture 2:} At large order of perturbation, the number of bubbles of a ``type 1'' graph scales like $|{V}_{\cal G}|/3$. This would in turn ensure that the degree of divergence of ``type 1'' graphs is uniform in the number of internal vertices.

Moreover, although not necessary from the point of view of field theory, we have also found the following natural conjecture to hold on all examples

{\bf Conjecture 3:} Type 1 graphs are manifolds of trivial topology, i-e isomorphic to the 3-sphere.

\medskip 

A final, technical point in this section is the following important property of ``type 1'' graphs, called the property {\it ``P''}.

\begin{definition}
 A graph is said to have the property ``P'' if for any two bubbles $b$ and $b'$ of ${\cal G}$,
 the graph $\bar{\cal G}^b\cap b'$ is such that {\bf any two faces} (that is closed strands) do {\bf not share} the same line.
\end{definition}

Before proceeding we will prove the lemma
\begin{lemma}\label{lemma:tech}
``Type 1'' graphs obey the property ``P''.
\end{lemma}
The proof is very simple and relies basically only on the analysis of figure \ref{fig:vertex1}.
What we will see in the following is that there exists graphs which satisfy the property P but are not of type 1 , so these two conditions are not equivalent.

{\bf Proof:}
By the first hypothesis of being of type 1, the 2D vertices $(1_2,1_3,1_4)$ and $(2_1,2_3,2_4)$ have to belong to two distinct bubbles, say $b$ and $b'$. They share the descendants $1_2$ and $2_1$ of the strand $(1,2)$.  Consequently $\bar{\cal G}^b\cap b'$ contains the whole face $2_1$ (and symmetrically $\bar{\cal G}^{b'}\cap b$ contains $1_2$).
On the contrary, the faces $1_3$ and $1_4$ are shared by $1$ and $3$ and $1$ and $4$. But each of  $1$, $2$, $3$, and $4$ belong to a different bubbles, as the graph is type 1. 
Hence the bubbles $b$ and $b'$ share just one face coming from the vertex (1234), and none of its neighbors. As this is true for any vertex, the two bubbles will share a set of faces not neighboring each other (that is, not sharing any lines).

\hfill $\Box$

\begin{figure}[hbt]
\centerline{
\includegraphics[width=60mm]{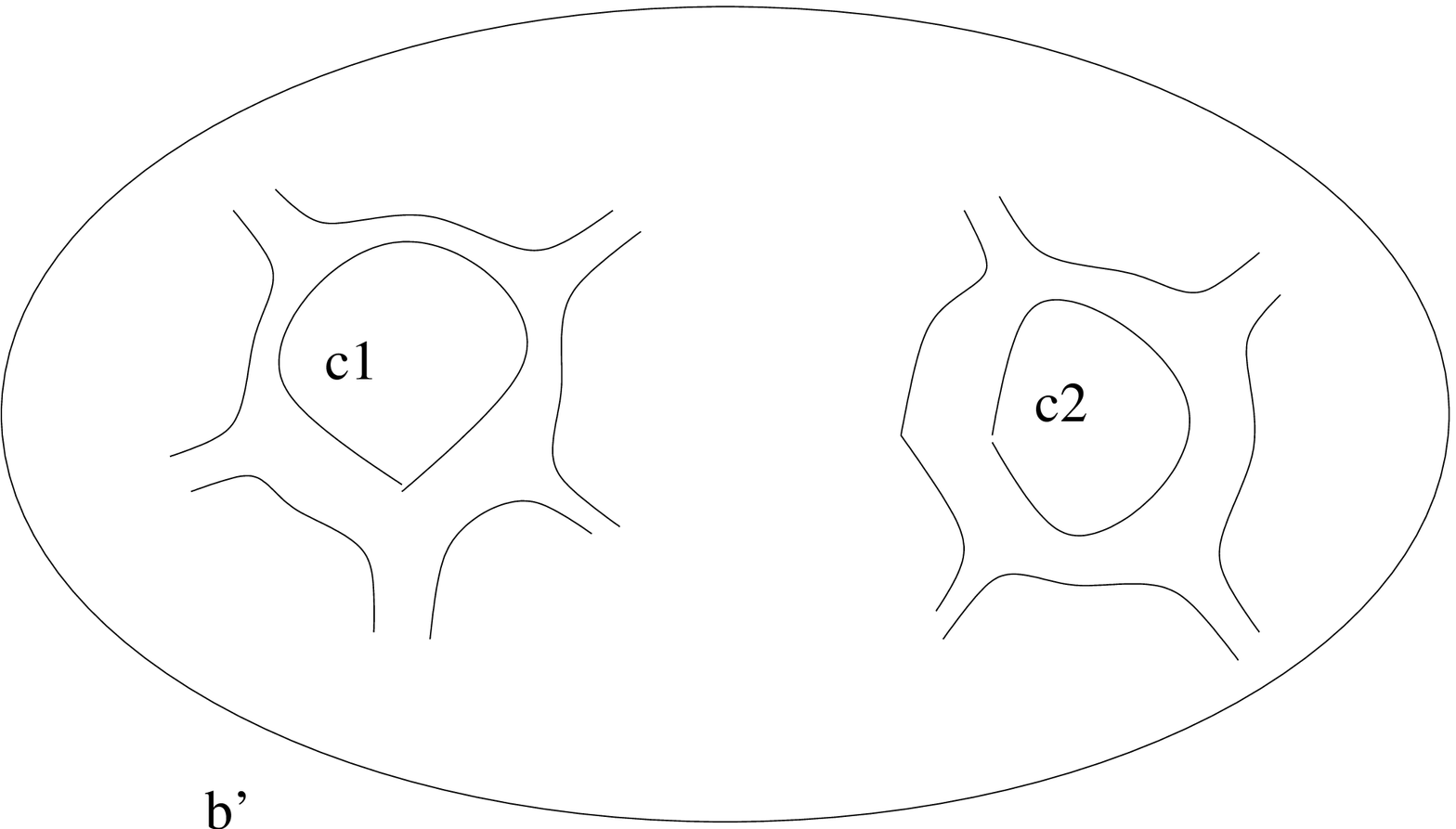}}
\caption{$\bar{\cal G}^b\cap b'$.}\label{fig:inter}
\end{figure}

This is represented in figure \ref{fig:inter}.

We will see later on that this property is a crucial ingredient of the contraction procedure we define for type 1 graphs, and thus for our main result concerning the power counting for the same class of graphs.

\section{Type 1 graphs in two dimensions}

Before addressing the three dimensional case we will detail the two dimensional ``type 1'' graphs. The importance of this study is twofold. First we will show that these graphs in two dimensions are exactly planar, one particle irreducible graphs, thus strengthening our claim that the ``type 1'' graphs are dominant in three dimensions. Second, the proof rely on a  contraction/deletion technique, similar to that of \cite{razvan1} which is then generalized and applied to for our results in three dimensions.

\begin{figure}[htb]
\centerline{
\includegraphics[width=100mm]{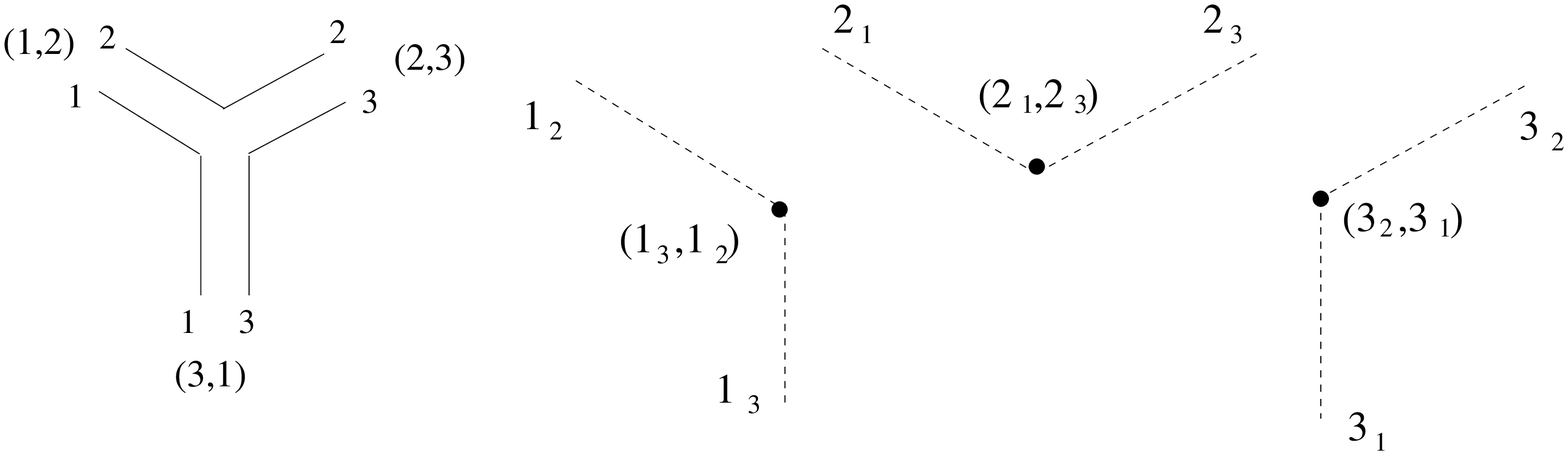}}
\caption{Two neighboring faces.}\label{fig:2D1Dproj}
\end{figure}

We start by generalizing the projection in a straightforward way to two-dimensional graphs, as shown in figure \ref{fig:2D1Dproj}.
The two-dimensional vertex $(1,2,3)$ projects into {\it three} one-dimensional vertices
$(1_3,1_2)$, $(2_1,2_3)$ and $(3_2,3_1)$, represented by dots in figure \ref{fig:2D1Dproj}. Each two-dimensional halfline projects into {\it two} one-dimensional halfline;
 $(1,2)$ for instance projects into $1_2$ and $2_1$. The two-dimensional bubbles are the faces of the graph, hence we denote them as $F\in{\cal F}$. Each of them projects into {\it one} one-dimensional connected component, $f$. Note that the 1D graphs (the connected components corresponding to the bubbles) obtained by projection are particularly trivial: they are cycles of lines and vertices of coordination two. The property ``P'', for instance, translates trivially for two dimensional graphs as
\begin{definition}
 A two dimensional graph $\cG$ is said to have the property ``P'' if for any two faces $f$ and 
$f'$ of $\cG$ the graph $\bcG^f\cap f'$ is such that any two { lines} do not share a { vertex}.
\end{definition}
Here we used again the notation $\bcG^f$ for the 1D projection of the 2D ancestor of the face $f$. One can also translate the definition of ``type 1'' graphs as

\begin{definition} \label{def:2Dtype1}
 A graph ${\cal G}$ is called ``type 1'' if
  \begin{itemize} 
 \item  $\forall \; f$, any 2D vertex of ${\cal G}^b$ (and consequently 2D line) projects 
into an { unique} 1D vertex (or line) of $f$.
  \item For all $f_1$, there exists an ordering $f_1<f_2\dots < f_{|{ F}_{\cal G}|}$ such that
\begin{itemize}
\item  $f_i \cap \bar{\cal G}^{f_1\cup \dots \cup f_{i-1}}$ is connected, for all$\;i \le |{ F}_{\cal G}|$.
\item $ f_i$ is not  contained  in   $\bar{\cal G}^{f_1 \cup f_2...\cup f_{i-1}}$ for $\;i < |{ B}_{\cal G}|$.
\end{itemize}
\end{itemize}
\end{definition}
Unsurprisingly, the ``type 1'' graphs have the property ``P''. The first condition of ``type 1'' implies that the three faces meeting at any two dimensional vertex are distinct, thus $1\neq2\neq3\neq1$ in figure \ref{fig:2D1Dproj}. Hence, any pair of neighboring lines like $(1,2)$ and $(1,3)$ can never be shared by the same two faces.

Generically, a 2D line $L$ separates two different faces. If one of these two faces is bounded {\it only} by the line $L$ (as it is for instance the case for a planar tadpole), we will call the latter a ``simple line''.

In 2D there exist two topological ``moves'', \cite{razvan1} which allow one to reduce a graph to an irreducible, topological equivalent one. These two moves are the contraction of a 2D tree line and the deletion of a 2D simple line. The topology of an orientable 2D graph is entirely encoded in its Euler character, $2-2g=V-L+F$. The deletion of a tree line reduces both the number of vertices and lines by 1, while the deletion of a simple line reduces the both the number of lines and faces by 1. The Euler character is thus invariant under this moves, and subsequently the amplitude (in the scaling limit) is unchanged. However, the valence of the vertices of the graph changes. The graphs obtained by applying this simplifications will no longer be dual to simplicial complexes, but to more general cellular complexes.

A natural question arises. Given a 2D graph $\cG$, what is the effect of the 2D topological moves on its 1D projection $\bcG$? 

In order to answer this question Lets consider  a 2D line $L$, connecting the two 2D vertices $V$ and $V'$ and separating two\footnote{These two faces are not necessarily distinct.} faces $F_1$ and $F_2$. The two faces $F_1$ and $F_2$ project into two connected components $f_1$ and $f_2$, the line $L$ projects into two 1D lines $l_1\in f_1$ and $l_2\in f_2$ and the vertices $V$ and $V'$ project into the 1D vertices $v_1, v_1'\in f_1$ and $v_2,v_2'\in f_2$. $l_1$ connects $v_1$ and $v_1'$ while $l_2$ connects $v_2$ and $v_2'$.

First, let $L$ be a 2D tree line, that is $V \neq V'$. Consequently $v_1\neq v_1'$ and 
$v_2\neq v_2'$. The contraction of the 2D line $L$ (and the gluing of vertices $V$ and $V'$) projects into the contraction of the line $l_1$ (and gluing of $v_1$ and $v_1'$) in $b_1$, accompanied  by the contraction of $l_2$ (and gluing of $v_2$ and $v_2'$) in $b_2$.
Now suppose that $L$ is a simple line. Take $f_1$ the connected component formed only by the line $l_1$, that is $v_1=v_1'$. The deletion of the line $L$ projects in the deletion of the 1D connected component $f_1$ accompanied by the {\it contraction} of the 1D line $l_2$ in $f_2$, (and the gluing of $v_2$ and $v_2'$).

We are now in the position to state the main result of this section
\begin{theorem}
 A 2D graph $\cG$ is type 1 if and only if it is a planar one particle irreducible graph.
\end{theorem}
{\bf Proof:}

The idea of the proof is simple. We will first show that the total order associated to $f_1$ specifies a sequence of topological moves which ultimately reduce the graph to a planar tadpole.
Conversely, for a planar graph one can always chose a sequence of moves which reduce it to a planar tadpole. This sequence of moves will in turn uniquely define a total order for its faces. The detailed (somewhat technical) proof is presented below.

{\bf Let $\cG$ be a ``type 1'' graph.} 

   Suppose that $\cG$ is a one particle reducible graph. The reducibility line $L$, connecting 
the vertices $V_1$ and $V_2$ separates the 2D graph into two connected components ${\cal C}_1\ni V_1$ and ${\cal C}_2\ni V_2$. The subgraph ${\cal C}_1$ would have only one face broken by $L$, say $F$. Consequently, the vertex $V_1$ would have two 1D descendents in the connected component $f$, the projection of $F$. As this violates the first condition in definition \ref{def:2Dtype1}, we conclude that $\cG$ is one particle irreducible.

   Now take the first connected component $f_1$ in the order specified in definition 
\ref{def:2Dtype1}. 
Denote the 1D vertices of $f_1$ by $v_1, \dots, v_p$. Each of them descends from different 
2D vertices $V_1,\dots V_p$ belonging to the 2D ancestor $F_1$ of $f_1$. 
Consider all lines save one of $f_1$, $l_1=(v_1,v_2), \dots l_{p-1}=(v_{p-1}, v_p)$, and there 2D ancestors, $L_1=(V_1,V_2), \dots L_{p-1}=(V_{p-1}, V_p)$. As the projection is one to one and $l_1,\dots l_{p-1}$ do not form a loop, $L_1,\dots L_{p-1}$ do not form a loop either. As such they are tree lines and can be contracted.
After contraction $f_1$ has become a 1D connected component with only one vertex and one 
line $l_p$. Its 2D ancestor $L_p$ is then simple, and can be deleted. This shows that we can contract and delete $f_{1}$

The contraction/deletion of the component $f_1$ {\it contracts} each $f_k \cap \bcG^{f_1}$ in 
a unique 1D vertex. In particular, in the component $f_2$, the connected subgraph $f_2 \cap \bcG^{f_1}$ has been contracted to a unique vertex. Thus, for $f_2$, each of its 1D vertices will still have an unique 2D ancestor.
The ancestor of a tree in $f_2$ is still a set of tree lines in $\cG$, and thus we can iterate the procedure. 
Ultimately we exhaust all but one 2D lines. The final graph will have one vertex and one loop line and consequently is a planar tadpole graph. Hence $\cG$ is a planar graph.

{\bf Let $\cG$ be a planar, 1PI graph.}

Suppose there exists a 2D line $L$, with both strands belonging to the same face. For instance, suppose that $1=2$ in figure \ref{fig:2D1Dproj}. As $\cG$ is 1PI, we can always contract a maximal tree\footnote{A tree connecting all vertices of the graph $\cG$} {\it not} containing $L$. After contraction, the graph $\cG$ becomes a rosette\footnote{A graph with only one 2D vertex.}. On this rosette the two sides of $L$ are the same face and there must exist some other line, $L'$ connecting them. The two lines $L$ and $L'$ generate a crossing on the rosette which contradicts the hypothesis that $\cG$ is a planar.

Thus, all 2D lines $L$ must separates two different faces. This implies that  for trivalent\footnote{This does not hold for four valent vertices as the reader can easily check.} vertices (see figure \ref{fig:2D1Dproj}) the three descendents of the vertex belong to the distinct connected components. Hence the first condition of definition \ref{def:2Dtype1} holds.

\begin{figure}[htb]
\centerline{
\includegraphics[width=100mm]{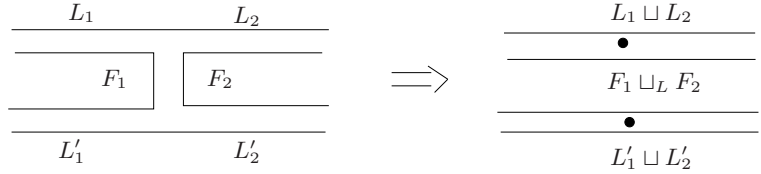}}
\caption{The joining of two faces separated by one line, $F_1\sqcup F_2$.}\label{fig:facesunion}
\end{figure}

We will now construct a total order on the connected components $f$ respecting the conditions stated in \ref{def:2Dtype1}. Chose any face, say $F_1$ and denote its projection $f_1$ as the first in the order. We will prove that, because $\cG$ is planar, there always exists a face, $F_2$ sharing exactly one line with $F_1$. We set its projection $f_2$ as the successor of $f_1$ in the total order. For two faces $F_1$ and $F_2$ sharing exactly one line $L=(V,V')$, we construct the face $(F_1 \sqcup_{L} F_2)$ by deleting the line $L$ {\bf and} the vertices $V, V'$, see figure \ref{fig:facesunion}. Consequently the lines $L_1,L_2$ and $L'_1,L'_2$ are joint into new lines $(L_1\sqcup L_2)$ and $(L'_1\sqcup L'_2)$.
In short we obtain by deleting the line L a new planar graph which is one particle irreducible, whose face are either initial face 
or the ``new face'' $(F_1 \sqcup_{L} F_2)$ and similarly edges are either initial edges or the new edges $(L_1\sqcup L_2)$ and $(L'_1\sqcup L'_2)$.
Since this graph satisfy our initial hypothesis we can find a face $F_{3}$ sharing only one edge with the face $(F_1 \sqcup_{L} F_2)$
and continue recursively the procedure till exhaustion of the faces.
By iteration we obtain a total order of the faces of the graph, respecting the requirements of definition \ref{def:2Dtype1}, thus proving that $\cG$ is a ``type 1'' graph.
\psfrag{lu}{$L_1$}
\psfrag{ld}{$L_2$}
\psfrag{fu}{$F_1$}
\psfrag{fd}{$F_2$}
\psfrag{lup}{$L'_1$}
\psfrag{ldp}{$L'_2$}
\psfrag{ll}{$L_1\sqcup L_2$}
\psfrag{ff}{$F_1\sqcup_{L} F_2$}
\psfrag{llp}{$L'_1\sqcup L'_2$}

\begin{figure}[htb]
\centerline{
\includegraphics[width=40mm]{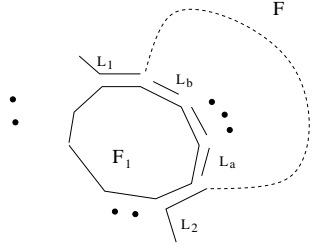}}
\caption{Two faces, $F_1$ and $F$, that share two lines in a planar graph.}\label{fig:faces}
\end{figure}

To conclude we just need to prove that for any face $F_1$ of a planar graph there exists a face $F_2$ sharing exactly one line with it. Let $F$ a face sharing at least a 2D line with $F_1$. 
Suppose $F_1$ and $F$ share two lines say at least two lines, we take $L_1$ and $L_2$ to be two consecutive 
such line as  in figure \ref{fig:faces}.
Note that $L_1$ and $L_2$ can not touch the same 2D vertex because the vertices are trivalent, and the three descendents of a vertex belong to different connected components. This shows that there exists lines $L_a, \dots$  between $L_1$ and $L_2$, encompassed by the face $F$ like in figure \ref{fig:faces}. Take the face $F_a$ which shares the line $L_a$ with $F_1$. If $F_{a}$ shares only one line then we are done. Otherwise $F_a$ and $F_1$ share two consecutive lines, $L_a$ and $L_b$. As the graph is planar, no line can intersect the dotted circuit in figure \ref{fig:faces}, hence $L_b$ must also be between $L_1$ and $L_2$. We the chose $F_a$ instead of $F$ and repeat the argument. We will always end up with a face $F_2$ which will share exactly one line with $F_1$.

\hfill $\Box$

\section{Examples of graphs}

In this section we will present some graphs 3D graphs, identify their 2D projections and compute their amplitudes. For simplicity we only consider vacuum graphs. 

\subsection{A ``type 1'' manifold graph.}
\label{sec:type1mani}
\begin{figure}[hbt]
\centerline{
\includegraphics[width=50mm]{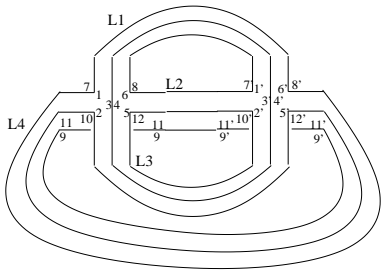}}
\caption{The sunshine graph ${\cal G}_1$.}\label{fig:sunmani}
\end{figure}

\begin{figure}[htb]
\centerline{
\includegraphics[width=80mm]{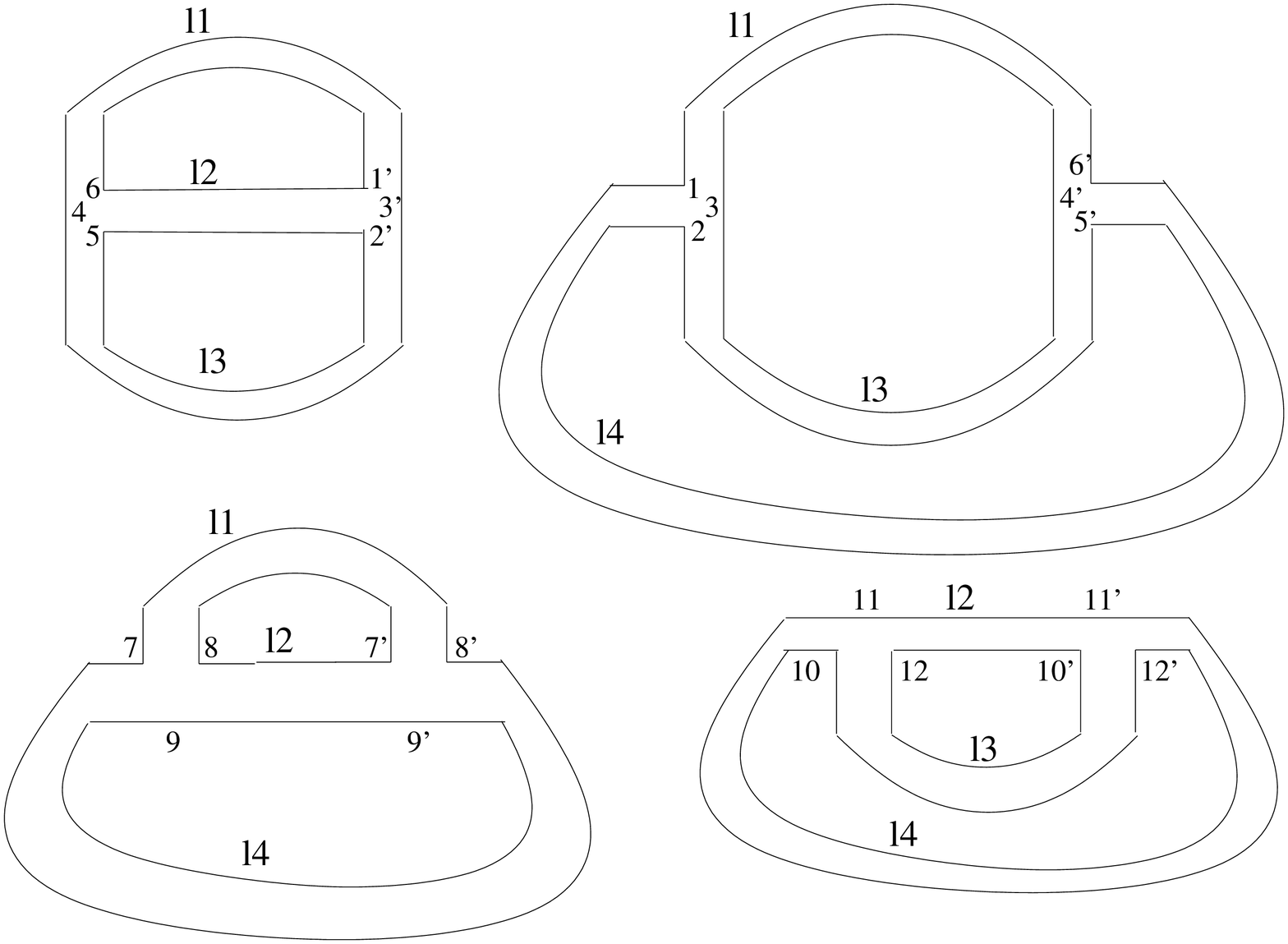}}
\caption{The 2D projection $\bar {\cal G}_1$.}\label{fig:bubsunmani}
\end{figure}

The first graphs, denoted ${\cal G}_1$ is drawn in figures \ref{fig:sunmani}. It has four bubbles, drawn in figure \ref{fig:bubsunmani}. We denote the four connected components of the 2D projection $\bar {\cal G}$ by $b_1=(l_1,l_2,l_3)$, $b_2=(l_1,l_3,l_4)$, $b_3=(l_1,l_2,l_4)$ and 
$b_4=(l_2,l_3,l_4)$\footnote{We do not distinguish between the 2D descendents of the same 3D line so not to make the drawings overly involved.}.
The relation between the topological numbers of the graph ${\cal G}_1$ and those of the graph $\bar {\cal G}_1$ can be directly verified. Thus, for instance, the 2D descendents of the 3D line $L_1$ belong to the components $b_1$ , $b_2$ and $b_3$ whereas the 2D descendants of the face $(L_1,L_2)$ belong to the $b_1$ and $b_3$.
In this particular case, all the bubbles $b_1$, $b_2$, $b_3$ and $b_4$ are planar (spheres), hence the graph ${\cal G}_1$ represents a manifold.
We can check for this particular graph that both conditions in definition \ref{def:type1} are fulfilled. Therefore ${\cal G}_1$ is a type 1 graph.

This graph is typical for the class of graphs for which we will establish the power counting in section \ref{sec:power}. Therefore, at the risk of being slightly pedantic, we will treat it in great detail in this section, so to give a good feeling of the general contraction procedure, and of the origin of the power counting result.
The task is, briefly, to identify some combinatorial substructure in the graph, 
which imply that the  corresponding group elements can be easily integrated out in evaluating the amplitudes.
Then we need to perform the relative integrations and obtain a general factorization property for the elements in the amplitude
 that correspond to the connected components of $\bar{\cal G}$, thus to the bubbles of the 3D graph ${\cal G}$, which is where the divergences of the model reside.   
We start by orienting all lines from the vertex $V=(1,\dots 12)$ to the vertex $V'=(1',\dots 12')$. Denoting the group elements associated to the 3D lines of the graph by $h_{L_1},\dots h_{L_4}$, the amplitude of ${\cal G}_1$ is
\bea \label{eq:ampl1}
A_{{\cal G}_1}&=&\int dh_{L_1} dh_{L_2}dh_{L_3}dh_{L_4} \; \delta^{\Lambda}(h_{L_1}h_{L_2}^{-1})\delta^{\Lambda}(h_{L_1}h_{L_3}^{-1})\delta^{\Lambda}(h_{L_1}h_{L_4}^{-1})
\delta^{\Lambda}(h_{L_2}h_{L_3}^{-1})\delta^{\Lambda}(h_{L_2}h_{L_4}^{-1})\delta^{\Lambda}(h_{L_3}h_{L_4}^{-1}) \;.
\eea

{\bf Step 1: Contraction of a tree in $b_1$.}
We choose the line $l_1$ which is a tree line in the component $b_1$ (actually, in this specific example, it corresponds to a full tree in $b_1$). Let $g_V$ and $g_{V'}$ two group elements, associated to the vertices $V$ and $V'$ respectively. We make the change of variables
\bea\label{eq:tree1}
h_{L_2}=g_V^{-1}h_{L_2}'g_{V'} \quad h_{L_3}=g_V^{-1}h_{L_3}'g_{V'} 
\quad  h_{L_4}=g_V^{-1}h_{L_4}'g_{V'} \quad h_{L_1}=g_V^{-1}g_{V'} 
\;.
\eea
Now we choose one group element $g_V$ among those appearing in the amplitude and we keep it fixed while computing the other integrations; this is the group element associated to the root of the tree we are considering. We get (neglecting the primes)
\bea \label{eq:ampli2}
A_{{\cal G}_1} &=&\int dg_{V'} dh_{L_2}dh_{L_3}dh_{L_4} \delta^{\Lambda}(h_{L_2}^{-1})\delta^{\Lambda}(h_{L_3}^{-1})
\delta^{\Lambda}(h_{L_4}^{-1})\,\delta^{\Lambda}(h_{L_2}h_{L_3}^{-1})\delta^{\Lambda}(h_{L_2}h_{L_4}^{-1})
\delta^{\Lambda}(h_{L_3}h_{L_4}^{-1})
\; .
\eea
We then see that the integral over $g_V$ gives $1$, as $g_V$ itself has disappeared from the amplitudes and the Haar measure is assumed to be normalized. 

\begin{figure}[htb]
\centerline{
\includegraphics[width=60mm]{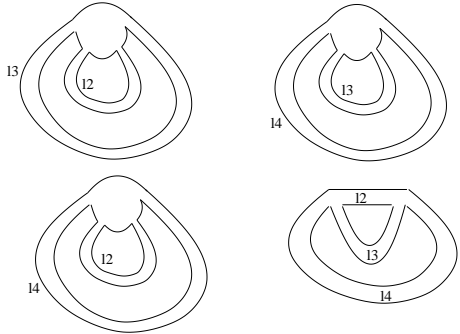}}
\caption{The graph $\bar {\cal G}'_1$.}\label{fig:bubredus}
\end{figure}

The remaining integrals in eq. \ref{eq:ampli2} correspond to a graph, (denoted ${\cal G}'_1$) obtained from ${\cal G}$ by contracting the line $L_1$. Consequently the 2D projection of ${\cal G}'_1$, denoted $\bar {\cal G}'_1$, is obtained from figure \ref{fig:bubsunmani} by contracting the lines $l_1$ (see figure \ref{fig:bubredus}).

We denote the connected components of $\bar {\cal G}'_1$ by $b'_1=(l_2,l_3)$, $b'_2=(l_3,l_4)$, $b'_3=(l_2,l_4)$ and 
$b'_4=(l_2,l_3,l_4)$. The component $b_1'$, obtained from $b_1$ by contracting $l_1$ is called a \lq\lq rosette\rq\rq, a graph topologically equivalent to $b_1$ but with only one vertex; the topological equivalence of the two graphs is easily verified as the contraction eliminates at once a line and a vertex, thus keeping the genus of the corresponding 2-surface invariant.

{\bf Step 2: Integration of all but one loop lines of $b_1$.}
All the delta functions in eq. (\ref{eq:ampli2}) correspond to faces of ${\cal G}'_1$, hence have two descendents in $\bar {\cal G}'_1$. Like the original graph, the two descendents of a 3D face belong to two distinct connected components, and any two connected components share the 2D descendents of at most one 3D face. This follows from the very definition of type 1 graphs. For instance, the 2D descendents of the 3D face $\delta(h_{L_2}^{-1})$ belong to $b_1'$ and $b_3'$ while the those corresponding to $\delta(h_{L_3}^{-1})$ belong to $b'_1$ and $b'_2$.

\begin{figure}[htb]
\centerline{
\includegraphics[width=60mm]{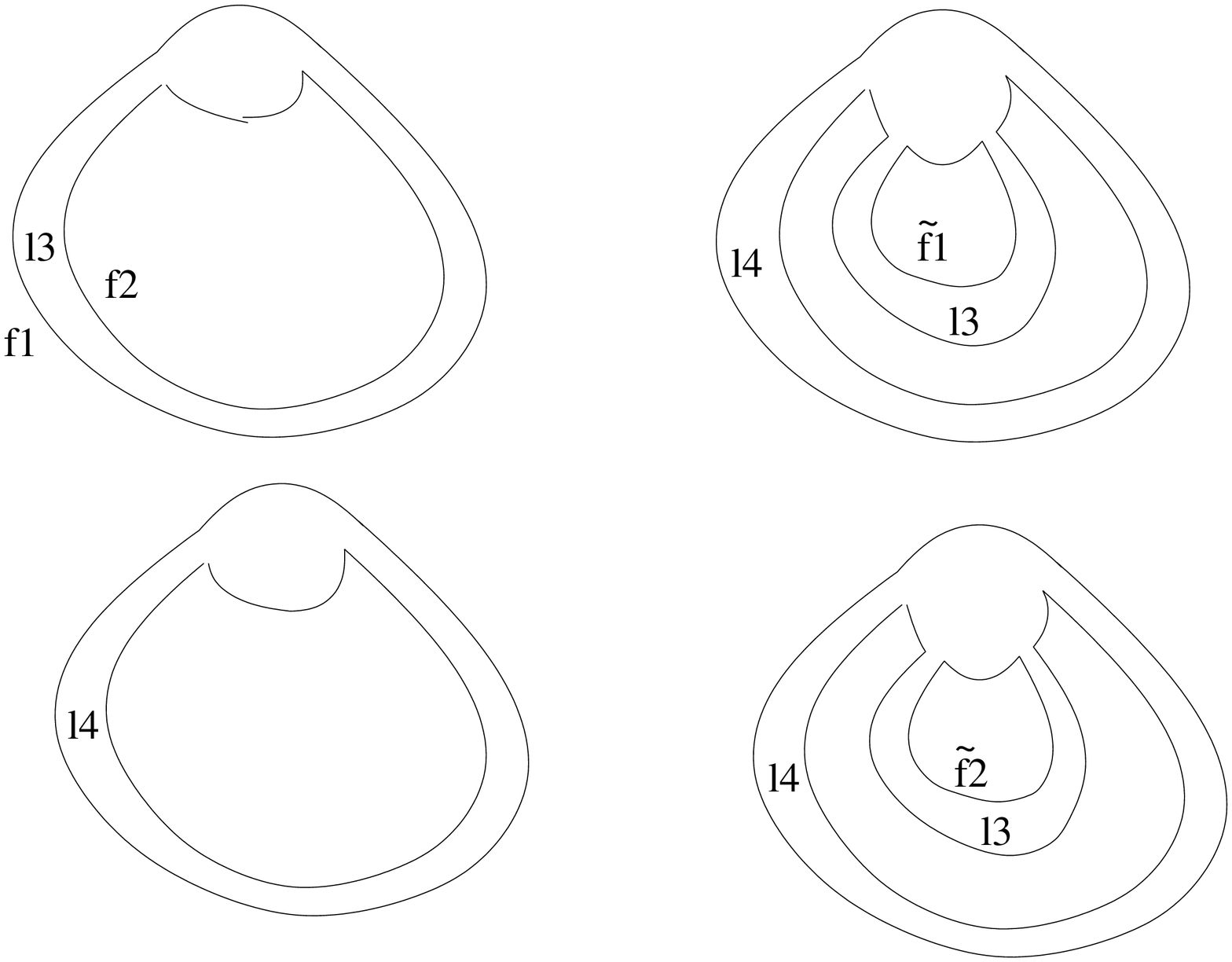}}
\caption{The graph $\bar {\cal G}''_1$.}\label{fig:bubredus1}
\end{figure}

$b_1'$ is a planar graph and has only one vertex. In consequence the loop lines do not cross. Consider the face bounded only by $l_2$ of $b_1'$. We integrate the group element $h_{L_2}$ using $\delta(h_{L_2}^{-1})$. Thus \ref{eq:ampli2} becomes
\bea\label{eq:ampli3}
A_{{\cal G}_1} &=&\int dh_{L_3}dh_{L_4} \delta^{\Lambda}(h_{L_3}^{-1})
\delta^{\Lambda}(h_{L_4}^{-1})
\delta^{\Lambda}(h_{L_3}^{-1})\delta^{\Lambda}(h_{L_4}^{-1})
\delta^{\Lambda}(h_{L_3}h_{L_4}^{-1})
\; .
\eea
The line $l_2$ appeared in three instances in the graph $\bar{\cal G}'_1$: twice as a loop line (on $b'_1$ and $b'_3$) and once as a tree line (on $b'_4$). The integration of $h_{L_2}$ has as consequence to delete $l_2$ each time it appeared as a loop line, but to {\bf contract} it where it appeared as a tree line! Thus the new expression \ref{eq:ampli3} corresponds to the reduced graph $\bar {\cal G}''_1$ presented in figure \ref{fig:bubredus1}. For more complicated graphs, we would repeat the procedure for all the lines forming our tree. 

{\bf Step 3: Factorization of $b_1$ using the last loop line.}
The connected components of ${\bar G}''$ are denoted $b_1'',\dots b_4''$. $b''_1$ has only one surviving loop line. It has two faces with amplitude $\delta(h_{L_3}^{-1})$. We call $f_1$ the exterior face and $f_2$ the interior one in figure \ref{fig:bubredus1}. The two faces can not be descendents of the same 3D face (as the descendents belong to different bubbles, because of the type 1 condition). In fact the copy of the interior face $f_2$ (called $\tilde f_2$) belongs to the bubble $b_4''$, while the copy of the exterior face $f_1$ (called $\tilde f_1$) belongs to $b_2''$.

We conclude that $\delta(h_{L_3}^{-1})$ appears twice in the expression of the amplitude (as can be checked directly on eq. \ref{eq:ampli3}). This implies that all the 2D descendents $l_3$ of $L_3$ are loop lines, belonging to different bubbles.

We integrate $h_{L_3}^{-1}$. The amplitude becomes
\bea\label{eq:ampli4}
A_{{\cal G}_1} &=&\delta^{\Lambda}(\id)\int dh_{L_4} \delta^{\Lambda}(h_{L_4}^{-1})
\delta^{\Lambda}(h_{L_4}^{-1})\delta^{\Lambda}(h_{L_4}^{-1})
\; .
\eea

\begin{figure}[htb]
\centerline{
\includegraphics[width=60mm]{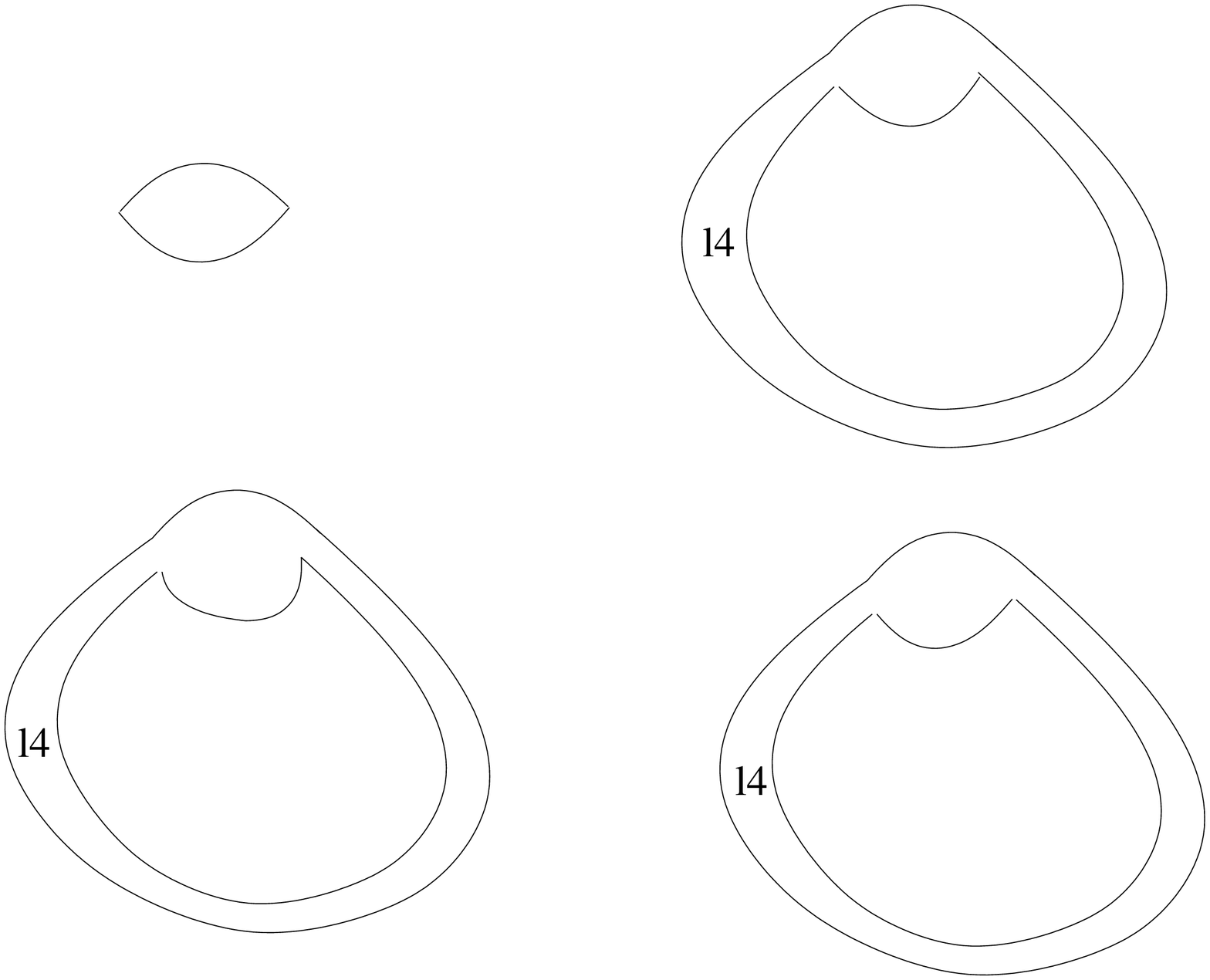}}
\caption{The graph $\bar {\cal G}'''_1$.}\label{fig:bubredus2}
\end{figure}

It corresponds to the graph $\bar {\cal G}'''_1$ obtained from $\bar {\cal G}''_1$ by deleting  all the 2D descendents $l_3$ of $L_3$ (see figure \ref{fig:bubredus2}).

We are now in the situation where the only surviving 2D lines $l_4$ are all descendents of the same 3D line $L_4$. They belong to three distinct bubbles. Integrating $h_{L_4}$ yields
\bea\label{eq:ampli7}
A_{{\cal G}_1} =[\delta^{\Lambda}(\id) ]^3.
\eea
This is the final result for the amplitude. It corresponds to having, in absence of the cut-off we imposed for regularizing the Feynman amplitudes of our model, one delta function divergence for each bubble (which has necessarily a boundary of spherical topology, as we assumed ${\cal G}$ to be a manifold) in the 3D graph ${\cal G}$, and with an amplitude that is completely factorized per bubble.

\subsection{A ``type 2'' manifold graph}
We now consider a similar graph, obtained from the previous one by a simple change in the gluing of the two GFT vertices it is composed of. The change amounts to an exchange between two of the lines of propagation, but is has the important consequence of violating the first type 1 condition: one of the faces in the projected graph $\bar{\cal G}$ in fact appears twice in the same bubble. 
As in this subsection and in the next we will simply compute the amplitudes without detailing the various steps of the computation, we will not use distinct letters for the 3D lines and there 2D descendents.

\begin{figure}[hbt]
\centerline{
\includegraphics[width=50mm]{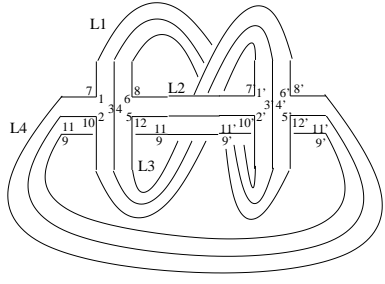}}
\caption{The twisted sunshine graph ${\cal G}_2$.}\label{fig:sunstrmani}
\end{figure}

\begin{figure}[hbt]
\centerline{
\includegraphics[width=100mm]{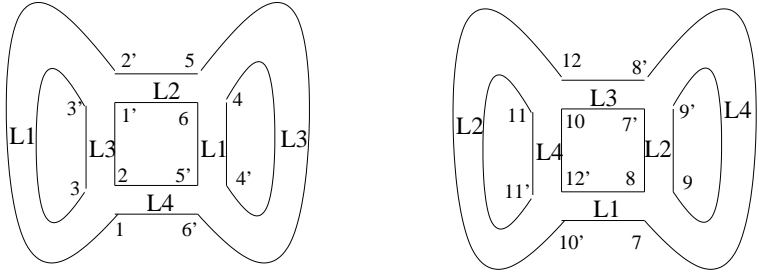}}
\caption{Bubbles of the graph ${\cal G}_2$ (the graph $\bar {\cal G}_2$).}\label{fig:bubsunstrmani}
\end{figure}

Consider now the graph ${\cal G}_2$, depicted in figure \ref{fig:sunstrmani}. Its bubbles are once more easily identified and are drawn in figure
\ref{fig:bubsunstrmani}.

First of all, notice that, as the boundaries of the two bubbles are both 2-spheres, this graph represents a manifold just as the previous example we considered. Unlike the previous example, however, as we noted, the face $(L_1, L_2)$ appears twice on the same bubble hence it is a type 2 graph. 
However as we will see this manifold is such that its first homotopy group is non trivial that is $\Pi_{1}(M)= \mathbb{Z}_{2}$,
it is therefore not a 3-sphere.
This is consistent with our conjecture that all type 1 graphs are manifolds of trivial topology.

The amplitude of this graph is given by:
\bea
A_{{\cal G}_2}&=&\int dh_{L_1}dh_{L_2}dh_{L_3}dh_{L_4}\; \delta^{\Lambda}(h_{L_1}h_{L_2}^{-1}h_{L_3} h_{L_4}^{-1})
\delta^{\Lambda}(h_{L_1}h_{L_3}^{-1})
\delta^{\Lambda}(h_{L_1}h_{L_4}^{-1}h_{L_3} h_{L_2}^{-1})\delta^{\Lambda}(h_{L_2}h_{L_4}^{-1})
\eea
and using the same change of variables as in eq. (\ref{eq:tree1}) it becomes
\bea
A_{{\cal G}_2}&=&\int dh_{L_1}dh_{L_2}dh_{L_3}dh_{L_4}\; \delta^{\Lambda}(h_{L_2}^{-1}h_{L_3} h_{L_4}^{-1})
\delta^{\Lambda}(h_{L_3}^{-1})\delta^{\Lambda}(h_{L_4}^{-1}h_{L_3} h_{L_2}^{-1})\delta^{\Lambda}(h_{L_2}h_{L_4}^{-1})\nonumber\\
&=&\int dh_{L_2} dh_{L_4}\; \delta^{\Lambda}(h_{L_2}^{-1} h_{L_4}^{-1})
\delta^{\Lambda}(h_{L_4}^{-1} h_{L_2}^{-1})\delta^{\Lambda}(h_{L_2}h_{L_4}^{-1})\nonumber\\
&=& \delta^{\Lambda}(\id)\int dh_{L_2}\; \delta^{\Lambda}(h_{L_2}^{-1} h_{L_2}^{-1}) \; .
\eea
where in the last equality we have given the dominant divergent contribution in the large $\Lambda$ limit.
The last integration can easily be performed in the $SO(3)$ case, for example, giving a divergent (in absence of regularization) contribution of the form $\sum_{n\in \mathbb{N}}  (2n+1)$. As we will see later on, the failure to satisfy the type 1 conditions prevents  the power counting formula we prove in this work to be satisfied.
 It shows that power counting is not uniform in the number of vertices so that, 
 if type 1 graphs can be made to dominate over type 2 graphs, this has to be in some scaling limit.
 One sees however that for a given order in perturbation theory the type 1 graph is the most divergent, having more bubbles.

\subsection{A \lq\lq type 2\rq\rq pseudomanifold graph}
\begin{figure}[hbt]
\centerline{
\includegraphics[width=50mm]{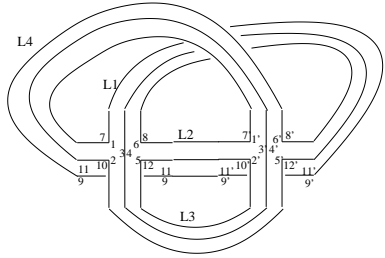}}
\caption{The permuted sunshine graph ${\cal G}_3$.}\label{fig:sunnonmani}
\end{figure}

\begin{figure}[hbt]
\centerline{
\includegraphics[width=100mm]{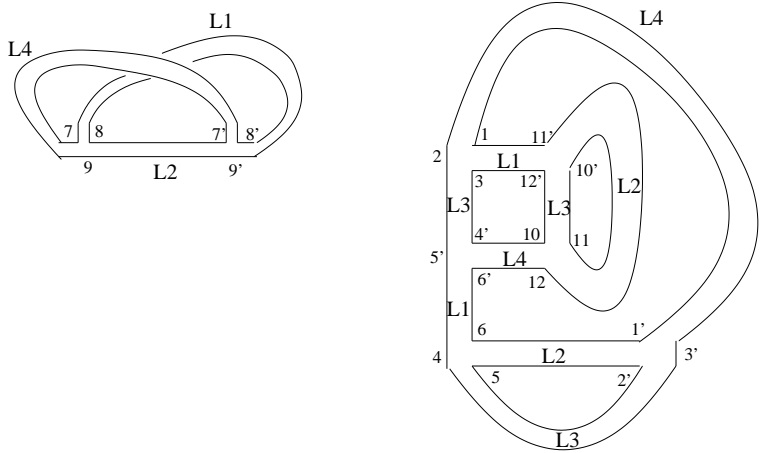}}
\caption{Connected components associated to the bubbles of the graph ${\cal G}_3$ (the $\bar {\cal G}_3$ graph).}\label{fig:bubsunnonmani}
\end{figure}

As our last example, we now consider another variation of the same 2-vertex graph. This time, we add a further exchange in the gluing of the two vertices. This has both the consequences that the resulting cellular complex fails to satisfy the type 1 conditions, and it also fails to satisfy the manifold conditions, since once of its bubble has the topology of a torus, as it can be easily checked.
 Moreover this graph possess two bubbles which share only one face thus the graph satisfies the property P even if it is not type 1.
 Our third graph, called ${\cal G}_3$ and its bubble $\bar {\cal G}_3$ are drawn in figures \ref{fig:sunnonmani} and \ref{fig:bubsunnonmani} respectively.

The amplitude associated to ${\cal G}_3$ is
\bea
A_{{\cal G}_3}&=&\int dh_{L_1}dh_{L_2}dh_{L_3}dh_{L_4}\; 
\delta^{\Lambda}(h_{L_1}h_{L_2}^{-1}h_{L_4}h_{L_1}^{-1}h_{L_2}h_{L_4}^{-1}) \delta^{\Lambda}(h_{L_1}h_{L_3}^{-1}h_{L_4}h_{L_3}^{-1})
\delta^{\Lambda}(h_{L_2}h_{L_3}^{-1})
\eea
which rewrites using again eq. (\ref{eq:tree1}) as
\bea
A_{{\cal G}_3}&=&\int dh_{L_2}dh_{L_3}dh_{L_4}\; 
\delta^{\Lambda}(h_{L_2}^{-1}h_{L_4}h_{L_2}h_{L_4}^{-1}) \delta^{\Lambda}(h_{L_3}^{-1}h_{L_4}h_{L_3}^{-1})
\delta^{\Lambda}(h_{L_2}h_{L_3}^{-1})\nonumber\\
&&\int dh_{L_2}dh_{L_4}
\delta^{\Lambda}(h_{L_2}^{-1}h_{L_4}h_{L_2}h_{L_4}^{-1}) \delta^{\Lambda}(h_{L_2}^{-1}h_{L_4}h_{L_2}^{-1}) \; .
\eea
The second delta function tells us that $h_{L_4}=h_{L_2}^2$ and substituting this in the first delta function yield
\bea
A_{{\cal G}_3}=\delta(\id) \int dh_{L_2}dh_{L_4}\delta^{\Lambda}(h_{L_2}^{-1}h_{L_4}h_{L_2}^{-1})=
\delta(\id)\; .
\eea
We thus see that this graph has a divergence (in absence of regularization) that is again of delta function type for each bubble of the graph minus one, as in the first example we have shown, and as our power counting theorem gives, in spite of the fact that, contrary to that example, it does not define a manifold. This implies that the topology of the cellular complex is reflected in the divergence of he amplitudes of the model in a way that is far from trivial. Also, this further example supports the conjecture that type 2 graphs are in general less divergent than type 1 graphs, at the same order in perturbation theory. This is further confirmed by all the examples we considered at higher (but still low) order.

\section{Power counting.}
\label{sec:power}
In this section we will prove a power counting theorem for ``type 1'' manifold graphs. The first part of our derivation can be applied, actually, to an arbitrary graph. It is only in the second part that the restriction to ``type 1'' manifolds will play a role.

The idea of the proof is simple. We apply topological moves to the given graph ${\cal G}$ to obtain from it a second graph ${\cal G}'$ , such that the quantum amplitudes of the two graphs are equal. Applying then a well defined sequence of further moves we simplify the graph ${\cal G}'$, and are in the end able to compute the amplitude. The point is that the topological moves used are devised in such a way that they not only simplify the graph, but also separate the contribution to the amplitude coming from the different bubbles. This sequence of topological moves is what we call a contraction process, which we highlight as the basis of the GFT renormalization procedure.

For simplicity we will restrict our attention to vacuum graphs. The generalization to graphs with arbitrary numbers of external legs is straightforward.

Before we proceed, we note here an important point of the construction. Our topological moves lead, from diagrams that are dual to simplicial complexes, i.e. the original GFT Feynman diagrams, to cellular complexes with more general combinatorics. Our construction therefore works and our theorem holds for a larger class of diagrams than those obtained from the Boulatov model. In other words, our results apply to any GFT model in 3 dimensions (i.e. with a field depending on three group arguments and still corresponding to a triangle) with trivial vertex and kinetic terms in the action (i.e. given by delta functions on the group), as in the Boulatov model, but with interaction terms of arbitrary order provided the identifications are still pairwise and the corresponding complexes are oriented (with outward triangle normals). Therefore, we consider a slightly larger category of graphs than those of the initial GFT, namely the vertices in our 3D graphs have arbitrary coordination. The dual of such a vertex is a convex polyhedron with the boundary triangulated by triangles duals to the half lines in the graph. Consequently the 2D graphs embedded in the 3D graph (the bubbles) will have vertices of arbitrary coordination as well. Notice that such GFT model will have again Feynman amplitudes given by delta functions over the group, whose arguments will again be holonomies of an $SU(2)$ connection around closed path encircling the 3D edges of a cellular complex; in other words, it will still correspond to a GFT quantization of 3D BF theory (and thus strictly related to 3D gravity), just like the simpler Boulatov model, now discretized on a more general cellular complex.  

\subsection{The Contraction of a Tree}
We prove first a lemma stating the invariance of the GFT amplitudes for the class of models we consider under part of the contraction procedure for Feynman graphs we define. 
\label{sec:3Dtree}
\begin{lemma}\label{lemma:tree}
 The amplitude of a graph is invariant under the contraction of a 3D tree line.
\end{lemma}
\noindent
{\bf Proof:}
Consider an arbitrary graph ${\cal G}$, and choose a tree line $L$ exiting from the vertex $V_p$ and entering in the vertex $V_q$. To this tree line, one single group element $h_L$ is associated, which will, by construction, be shared by three (not necessarily distinct) faces of the graph. We denote by  $L_p\in V_p$ ($L_q\in V_q$) all other lines of ${\cal G}$ touching $V_p$ ($V_q$). We now want to isolate and characterize, first, the contribution of this tree line to the full amplitude, and then check the effect of removing this line from the graph, by integrating it out. 
Let us focus on the faces of the graph. Looking at their relation with the tree line $L$, they can be classified in five distinct types:
\begin{itemize}
\item The face of type $F_0$ does not touch neither of the vertices $V_p$ or $V_q$.
\item The face of type $F_1$ goes from the vertex $V_p$, through the line $L$, to the vertex $V_q$. 
\item The face of type $F_2$ goes from the vertex $V_q$, through the line $L$, to the vertex $V_p$.
\item The face of type $F_3$ touches the vertex $V_p$ with two strands but does not cross over to the vertex $V_q$; therefore its boundary does not include the tree line $L$.
\item The face of type $F_4$ touches the vertex $V_q$ with two strands but does not cross over to the vertex $V_p$; therefore its boundary does not include the tree line $L$.
\end{itemize}
Clearly, the three faces whose boundary includes the tree line $L$ can only be of type $F_1$ or $F_2$, and the delta functions associated to these faces will of course contain the group element $h_L$, or its inverse $h_L^{-1}$, respectively, in their argument; so we can have three $F_1$ faces, three $F_2$ faces or two faces of one type and one of the other. The effect of the contraction of the tree line $L$ is different for the above five types of faces, and this is what we need to study here.
We can highlight the contribution of the vertices $V_p$ and $V_q$, touched by the tree line $L$, to the full amplitude $A_{\cal G}$ by writing the same amplitude as:
\bea
A_{\cal G} = \left( \prod_{l} \int dh_l\right) \prod_{F_0} \delta_{F_0}( ...h_l... )\, I(h_l), \nonumber
\eea
where $l$ labels the lines of ${\cal G}$ that do not touch the vertices $V_p$ and $V_q$, $F_0$ indicates the faces that only have such lines in their boundary, and $I$ represents instead the contribution to the amplitude coming from the faces that touch the vertices $V_q$ and $V_p$. This, we repeat, will be given by the three deltas associated to the three faces of type $F_1$ or $F_2$ and depending on $h_L$, times all the other deltas of type $F_3$ or $F_4$; all these deltas will a priori depend also on the group elements $h_l$.
In order to simplify the exposition, we assume that at least one of the three faces depending on $h_L$ is of type $F_1$ and at least one of them is of type $F_2$, we focus on these two faces and leave implicit the third one, because it will behave (under contraction of the tree line $L$), just as one of the other two. Similarly, we focus only on two of the faces of type $F_3$ and $F_4$, assuming that $I$ contains at least one face of each type. Again, this simplifies the presentation and makes the proof of the lemma clearer, but the same reasoning would hold in the most general case as well. 
The contribution of $L$ to the amplitude then writes:
\bea\label{eq:tream1}
I(h_l)&=&\left( \prod _{L_p \in V_p}\int dh_{L_p} \prod_{L_q\in V_q}\int dh_{L_q} \int dh_L\right) \, \delta_{F_1}(\dots h^{\sigma(L_{p_1})}_{L_{p_1}}h_Lh^{\sigma(L_{q_1})}_{L_{q_1}} \dots ) 
\delta_{F_2}(\dots h^{\sigma(L_{q_2})}_{L_{q_2}}h^{-1}_Lh^{\sigma(L_{p_2})}_{L_{p_2}} \dots )  \nonumber\\
&&
\delta_{F_3} (\dots h^{\sigma(L_{p_3})}_{L_{p_3}} h^{\sigma(L_{p_4})}_{L_{p_4}} \dots)\,\delta_{F_4} (\dots h^{\sigma(L_{q_3})}_{L_{q_3}}h^{\sigma(L_{q_4})}_{L_{q_4}}\dots )\,\prod_{F} \delta_{F}(....) \; ,
\eea
where $F$ labels the other face of type $F_1$ or $F_2$ that we know is contributing to the amplitude, and all the other faces of type $F_3$ and $F_4$ that we do not focus on.
The exponent $\sigma(L_{p_1})$ in $\delta_{F_1}$ is $1$ if the orientation of the line $L_p$ coincides with that of the face $F_1$ and $-1$ if not (and similarly for all other exponents).

Let us now associate two group elements $g_{V_p}$ and $g_{V_q}$  to the vertices $V_p$ and $V_q$, and change variables as follows
\bea
&&h_{L_p}=h_{L_p}'g_{V_p} \text{ if } h_{L_p} \text{ enters in the vertex } V_p\nonumber\\
&&h_{L_p}=g_{V_p}^{-1}h_{L_p}' \text{ if } h_{L_p} \text{ exits from the vertex } V_p \; ,
\eea
and similarly for $h_{L_q}$.

From eq. (\ref{eq:tream1}) we see that the face $F_1$ is oriented from the vertex $V_p$ to $V_q$. Then
\bea \label{eq:F1}
h^{\sigma(L_{p_1})}_{L_{p_1}}=h'^{\sigma(L_{p_1})}_{L_{p_1}} g_{V_p} \quad h^{\sigma(L_{q_1})}_{L_{q_1}}=g_{V_{q_1}}^{-1} h'^{\sigma(L_{q_1})}_{L_{q_1}}\; .
\eea

On the contrary the face $F_2$ is oriented from $V_q$ to $V_p$ and we have
\bea \label{eq:F2}
h^{\sigma(L_{p_2})}_{L_{p_2}}=g_{V_p}^{-1} h'^{\sigma(L_{p_2})}_{L_{p_2}} \quad h^{\sigma(L_{q_2})}_{L_{q_2}}= h'^{\sigma(L_{q_2})}_{L_{q_2}} g_{V_{q}}^{-1}\; .
\eea

The face $F_3$ is oriented from the line $L_{p_3}$ to $L_{p_4}$ and we have
\bea\label{eq:F3}
h^{\sigma(L_{p_3})}_{L_{p_3}}=h'^{\sigma(L_{p_1})}_{L_{p_1}} g_{V_p} \quad 
h^{\sigma(L_{p_4})}_{L_{p_4}}=g_{V_p}^{-1} h'^{\sigma(L_{p_4})}_{L_{p_4}} \; ,
\eea
and analogously for $F_4$.

We now fix $g_{V_q}$ fixed and change variable from $h_L$ to $g_{V_p}=g_{V_q} h^{-1}_L$. Using eq. \ref{eq:F1}, 
\ref{eq:F2} and \ref{eq:F3}, the amplitude \ref{eq:tream1} becomes
\bea\label{eq:tream2}
I(h_l)&=&\int \prod _{L_p \in V_p}dh'_{L_p} \prod_{L_q\in V_q}dh'_{L_q} dg_{V_p} \prod \delta_{F_1}(\dots h'^{\sigma(L_{p_1})}_{L_{p_1}}h'^{\sigma(L_{q_1})}_{L_{q_1}} \dots ) 
\delta_{F_2}(\dots h'^{\sigma(L_{q_2})}_{L_{q_2}}h'^{\sigma(L_{p_2})}_{L_{p_2}} \dots )\nonumber\\
&& 
\delta_{F_3} (\dots h'^{\sigma(L_{p_3})}_{L_{p_3}} h'^{\sigma(L_{p_4})}_{L_{p_4}} \dots) \delta_{F_4} (\dots h'^{\sigma(L_{q_3})}_{L_{q_3}}h'^{\sigma(L_{q_4})}_{L_{q_4}}\dots )\;\prod_{F} \delta_{F}(....) \; .
\eea

The integral over $g_{V_p}$ decouples and computes to one. The new amplitude corresponds to a graph with the same faces, same connectivity but where the line $L$ disappeared. We identify this graph as the one obtained from ${\cal G}$ with the two vertices $V_p$ and $V_q$ contracted using the tree line $L$.
\hfill $\square$

\medskip

We conclude this preliminary step toward proving our main result with an observation. Note that the descendents of any tree line in the 3D graph are tree lines in the 2D graph of bubbles, because different 3D vertices always project into different 2D vertices. Hence, the number of tree lines in the 2D graph must be at least be equal to the number of tree lines which descend from the 3D tree. This allows us to give an upper bound on the number of bubbles of an arbitrary graph:
\bea
3(|{V}_{\cal G}|-1)\le |{v}_{\cal G}|-|{b}_{\cal G}| \Rightarrow
|{B}_{\cal G}|\le |{V}_{\cal G}|+3\;.
\eea

The maximal number of bubbles decreases when taking into account the loop lines, hence improving the above bound.

\subsection{Amplitudes and power counting for ``type 1'' manifolds}
\setcounter{theorem}{0}
\begin{theorem}
 The amplitude of a connected ``type 1'' {\it manifold} vacuum 3D graph ${\cal G}$ is
   \bea
    A_{{\cal G}}\,=\,\left( \delta^{\Lambda}(\id)\right)^{|{B}_{\cal G}|-1}
   \eea
\end{theorem}
\noindent
{\bf Proof:} The proof is a straightforward generalization of the computation in section \ref{sec:type1mani}. Before we proceed we need to understand better the structure of a ``type 1'' graph.

In figure \ref{fig:propag} we depicted a 3D line $L=(a',b',c')\rightarrow (a,b,c)$, and its three 2D descendents $l^{b_1}= (c'_{b'},c'_{a'})\rightarrow (c_b,c_a)$,
$l^{b_2}= (b'_{a'},b'_{c'})\rightarrow (b_a,b_c)$ and $l^{b_3}= (a'_{c'},a'_{b'})\rightarrow (a_c,a_b)$, where the upper index of $l$ refers to the connected component to which the 2D line belongs. As the graph is ``type 1'', the three connected components $b_1$, $b_2$, and $b_3$ are all different.

Also from figure \ref{fig:propag}, we see that $b_1$ and $b_2$ share the two descendents of the same 3D face (the 2D strands $(c'_{b'},c_b)$ and $(b'_{c'},b_c)$ descend from the same 3D strand $(b',c')\rightarrow (b,c)$).

\begin{figure}[hbt]
\centerline{
\includegraphics[width=80mm]{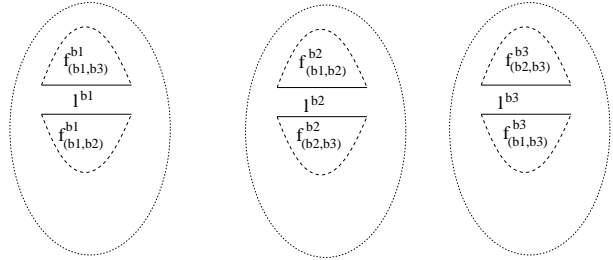}}
\caption{The three 2D descendents of the same 3D ancestor.}\label{fig:ty}
\end{figure}

We denote the face $(c'_{a'},c_a)$ by $f^{b_1}_{(b_{1},b_{2})}$, where the upper index denotes again the connected component to which the face belongs, and the lower indices denote the two connected components which share the two descendents of the 3D ancestor of the face. The same goes for the other couples of bubbles, $b_1,b_3$ and $b_2,b_3$. The situation and notations are then summarized in figure \ref{fig:ty}.

We now procede to the computation of the amplitude of a type 1 graph.

\medskip

{\bf Step 1: Contraction of a tree in a connected component $b'$.}

Consider a connected component $b'$ in $\bar {\cal G}$. 
Denote $v^{b'}$, $l^{b'}$, and $f^{b'}$ the sets of vertices 
lines and faces of $b'$. $b'$ is a planar connected 2D graph.
As ${\cal G}$ is ``type 1'', all the 2D vertices $v^{b'}$ are descendents of different 3D vertices $V^{b'}$ of ${\cal G}$ and all 2D lines $l^{b'}$ are descendants of different 3D lines $L^{b'}$.

Let $\mathfrak{t}^{b'}$ a set of $|v^{b'}|-1 $ lines forming a tree in the connected component $b'$. We denote ${\cal T}^{b'}$ the 3D ancestor of this set in ${\cal G}$.
We prove now that ${\cal T}^{b'}$ does not form any loops in the graph ${\cal G}$, i.e. it is itself a 3D tree; in general, it would not be a maximal tree though, but simply a set of tree lines; this can be easily checked by looking at specific examples. This fact is not trivial and in fact depends on the ``type 1'' condition for ${\cal G}$. So it is at this point that the assumption that our graph is type 1 plays a crucial role. 

The proof goes as follows. Suppose ${\cal T}^{b'}$ has a loop in ${\cal G}$. Let $V_1,\dots V_k$ and $L_1,\dots L_k$ the 3D vertices and lines belonging to this loop. $V_1$ is connected by $L_1$ with $V_2$, which is connected by $L_2$ to $V_3$, and so on until we reach $V_k$ which is connected by $L_k$ with $V_1$. Because ${\cal G}$ is ``type 1'', $L_1$ has a unique descendent on $b'$, called $l_1$, and $V_1$ and $V_2$ have unique descendents on $b'$, called $v_1$ and $v_2$.
$l_1$ necessarily connects  $v_1$ and $v_2$. Iterating we find that $l_1,\dots, l_k$ forms a loop in $b'$, which contradicts the hypothesis that $\mathfrak{t}^{b'}$ is a tree\footnote{If, on the contrary a vertex, let's say $V_2$, had 2 descendents on $b'$, then $l_1$ and $l_2$ could connect each to a different descendent.  Then the loop in ${\cal G}$ does not need to project into a loop in $b'$! This is why we need the first condition in the definition of ``type 1'' graphs.}.

Therefore, ${\cal T}^{b'}$ is a set of 3D tree lines, as anticipated. We now proceed by contracting all 3D lines in this set using the procedure described in section \ref{sec:3Dtree}. By lemma \ref{lemma:tree} the amplitudes are invariant under the contraction. The result from the 2D perspective is to contract {\it all} the 2D descendents of the tree lines in ${\cal T}^{b'}$.

Thus the 2D graph $b'$ reduces to a planar rosette, called $b$. We now focus on this 2D structure.

\medskip

{\bf Step 2: Deletion of the lines of $b$.}

Consider a line $l^b$ on the boundary of the bubble $b$. It bounds two faces $f^b_{(b,b_1)}$ and $f^b_{(b,b_2)}$. The 3D ancestor $L^b$ of $l^b$ has another two descendents. The first is $l^{b_1}\in b_1$ bounding $f^{b_1}_{(b_1,b)}$ and $f^{b_1}_{(b_1,b_2)}$ and the second is $l^{b_2}\in b_2$ bounding 
$f^{b_2}_{(b_2,b)}$ and $f^{b_2}_{(b_2,b_1)}$.

As the component $b$ is nonempty, there exist at least one face of $b$ bounded by only one line. Such a face exists, as the rosette $b$ is planar, thus having only lines which do not cross. We call such a face ``simple''.

Let $f^b_{(b,b_1)}$ be a simple face, bounded only by the line $l^b$. Then $f^{b_1}_{(b_1,b)}$ is also simple as the two faces share the same 3D ancestor.
We necessarily fall in one of the following four cases:
\begin{itemize}
 \item $f^{b}_{(b,b_2)}$ is not simple and $f^{b_1}_{(b_1,b_2)}$ is also not simple
 \item $f^{b}_{(b,b_2)}$ is not simple, but  $f^{b_1}_{(b_1,b_2)}$ is simple
 \item $f^{b}_{(b,b_2)}$ is simple, but $f^{b_1}_{(b_1,b_2)}$ is not simple
 \item both $f^{b}_{(b,b_2)}$ and $f^{b_1}_{(b_1,b_2)}$ are simple
\end{itemize}

\medskip

\begin{figure}[hbt]
\centerline{
\includegraphics[width=80mm]{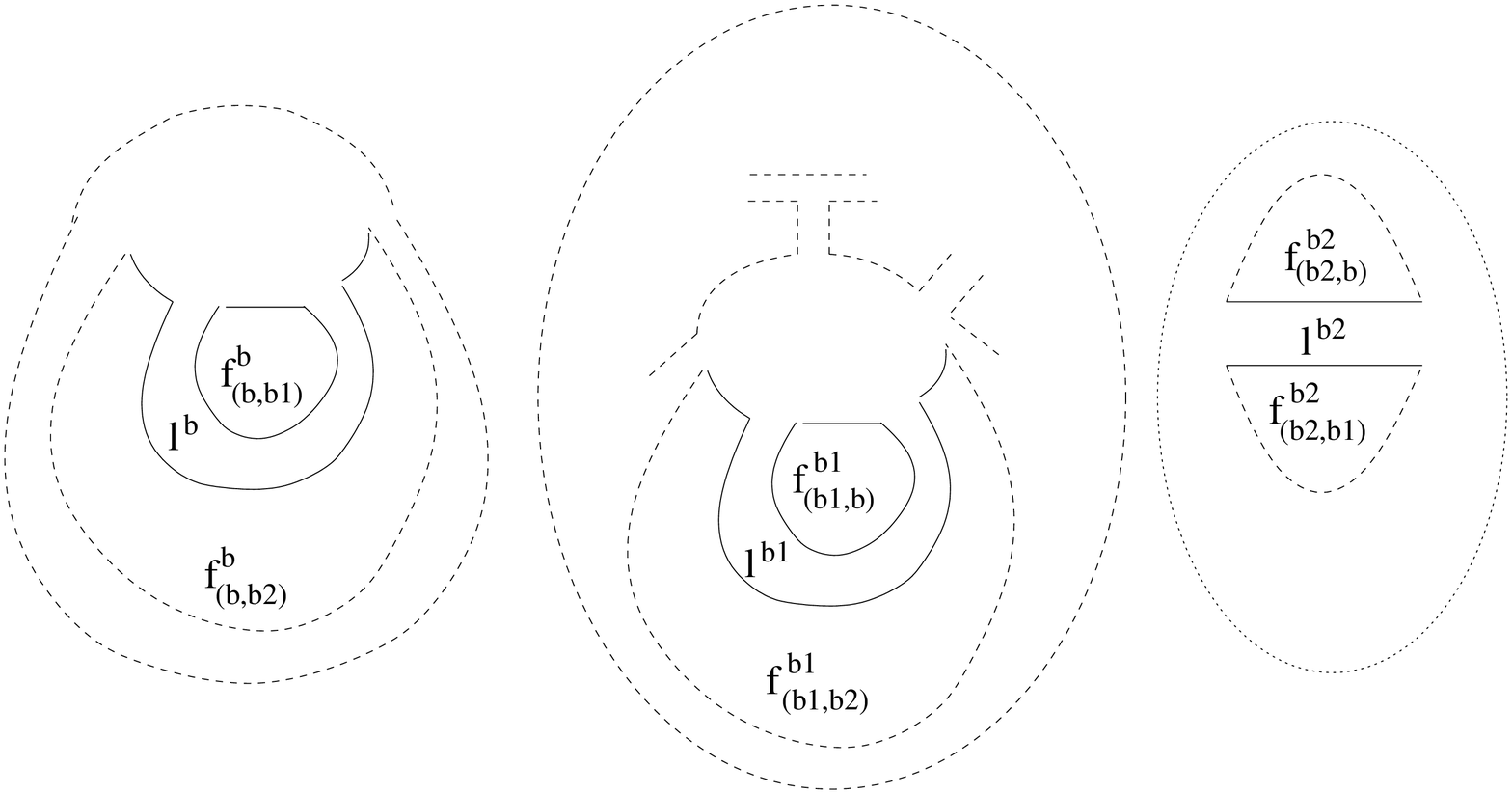}}
\caption{The first case.}\label{fig:caz1}
\end{figure}

{\bf Case 1} is depicted in figure \ref{fig:caz1}.

The integral corresponding to $h_{L^b}$ is
\bea
\int dh_{L^b} \delta(h_{L^b}) \delta(h_{L^b} f) \delta(h_{L^b} g)=
\delta(f) \delta(g) \; ,
\eea
where $f$ and $g$ are nontrivial group elements associated to the 3D ancestors of the two faces $f^{b}_{(b,b_2)}$ and $f^{b_1}_{(b_1,b_2)}$.

The graph ${\cal G}$ has the property ``P'', namely $\bar{\cal G}^{b_1}\cap b_i$ is such that no two faces share a line. Then, if the internal face of some connected component is not simple, the connected component has at least two 
vertices\footnote{As any two vertices belonging to this face can be contracted {\bf only} by propagators belonging to the face.}.
This implies that $l^{b_2}$ is necessarily a {\bf tree} line on the bubble $b_2$, as the face $f^{b_2}_{(b_2,b)}$ is not simple.

The consequence of this integration is that the lines $l^b$ and $l^{b_1}$ are deleted and the line $l^{b_2}$ is {\bf contracted}, as it is a tree line in $b_2$ (see figure \ref{fig:caz1}). As long as we remain in this case we proceed by iterating Step 2.

\medskip

\begin{figure}[hbt]
\centerline{
\includegraphics[width=80mm]{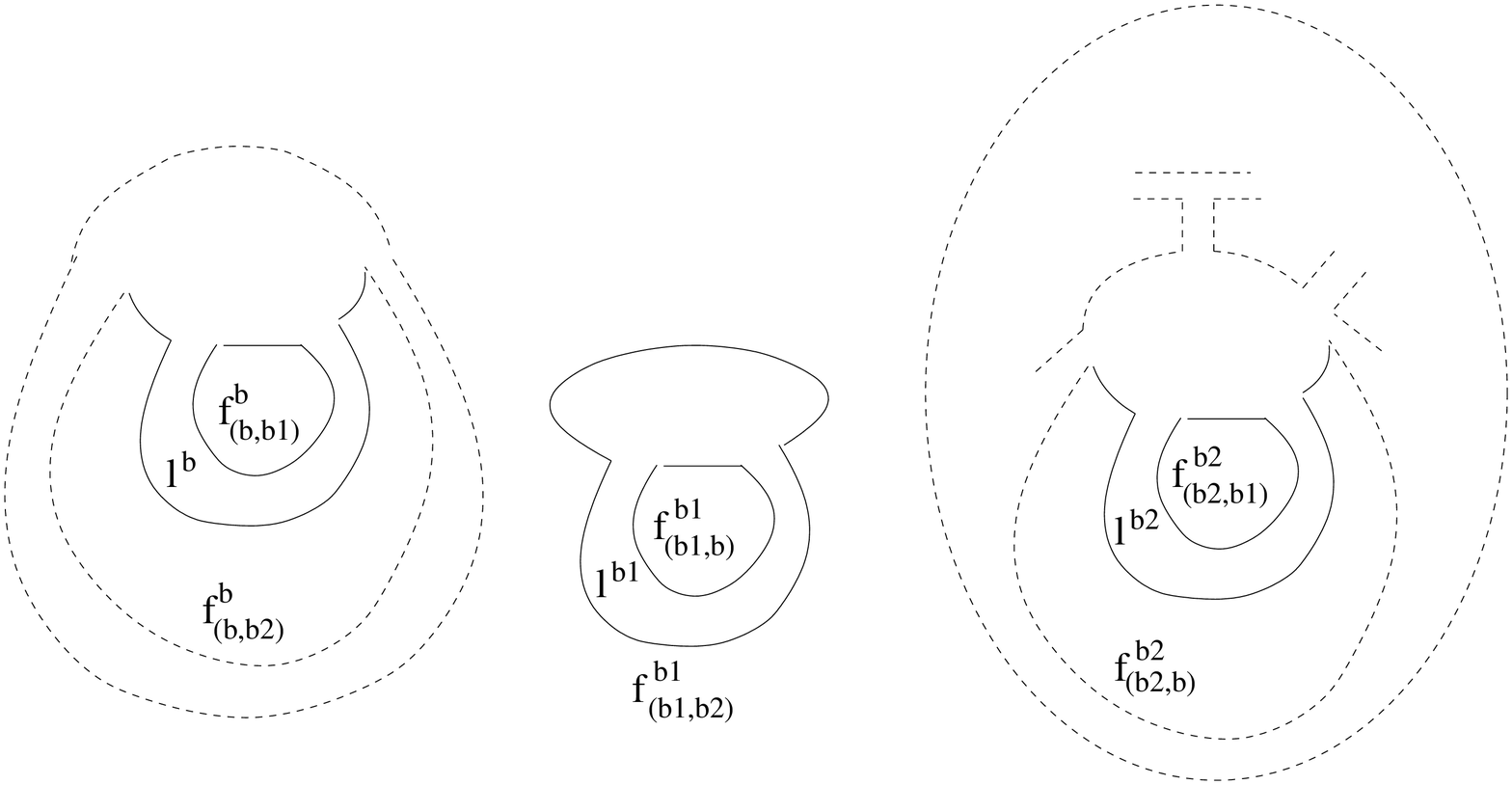}}
\caption{The second case.}\label{fig:caz2}
\end{figure}

{\bf Case 2} is depicted in figure \ref{fig:caz2}.

This is actually excluded for a ``type 1'' graph as  $b_1\setminus \bar{\cal G}^{b}$ contains at least a line and thus $f^{b_1}_{(b_1,b)}$ and $f^{b_1}_{(b_1,b_2)}$ can not both be simple at the same time.

\medskip

\begin{figure}[hbt]
\centerline{
\includegraphics[width=80mm]{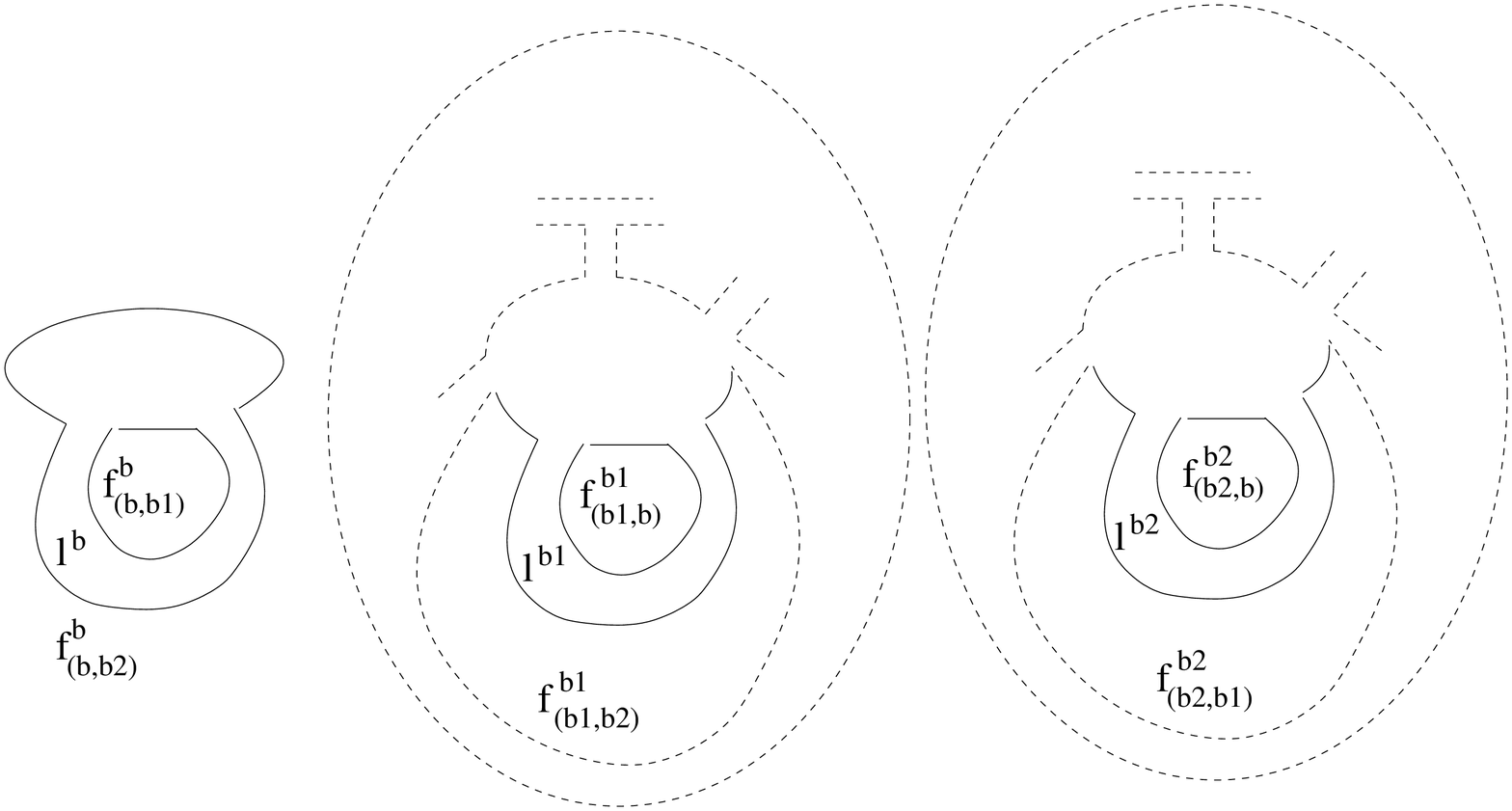}}
\caption{The third case.}\label{fig:caz3}
\end{figure}

{\bf Case 3}  is depicted in figure \ref{fig:caz3}.

The integral corresponding to $h_{L^b}$ is
\bea
\int dh_{L^b}\, \delta(h_{L^b}) \delta(h_{L^b}) \delta(h_{L^b} g)=
\delta(\id) \delta(g) \; ,
\eea
where $g$ is the nontrivial group elements associated to the 3D ancestor of the faces $f^{b_1}_{(b_1,b_2)}$.
In this case the lines $l^b, l^{b_1}$ and $l^{b_2}$ are all loop lines in the bubbles $b,b_1$
and $b_2$. 

The consequence of the integration is that they are all deleted.  Moreover, the bubble $b$ disconnects: it has no remaining lines (see figure \ref{fig:caz3}). This lead to a global factor $\delta(\id)$ in the amplitude.

\medskip

{\bf Iteration:} The contraction of a tree in $b_1$ shrinks its 3D ancestor ${\cal G}^{b_1}$ to a graph having an  unique 3D vertex (of arbitrary coordination), call it $V_b$, and some 3D tadpole lines (that is 3D lines with which start and end on the vertex $V_b$). The subsequent deletion of all loop lines of $b_1$ further simplifies this graph (that is, it eliminates the tadpole lines which have some descendant belonging to $b$).

As the graph is ``type 1'', $b_2\cap \bar{\cal G}^{b_1}$ is connected. Consequently the factoring out of $b_1$ contracts this connected subgraph graph to a 2D vertex, say $v_b$. Thus the 3D vertex $V_b$ projects on the bubble
$b_2$ into an {\bf unique} vertex $v_b$. This allows us to iterate the step 1, as the ancestor of a 2D tree on $b_2$ is a set of 3D tree lines.

\begin{figure}[hbt]
\centerline{
\includegraphics[width=60mm]{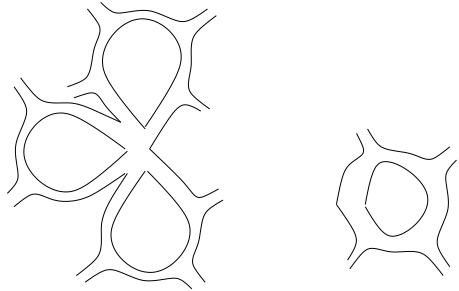}}
\caption{ $\bar{ {\cal G}^{b_1} } \cap b_j$ after the contraction of $b_1$.}\label{fig:intersection}
\end{figure}

Moreover, after factoring out the fist bubble $b_1$, all the subgraphs $\bar{\cal G}^{b _1} \cap b_k$ are fully contracted to points. Due to lemma \ref{lemma:tech} we see that $\bar {\cal G}^{b_2}  \cap b_j$ is a set of faces such that no two of them share a line (then typically a situation like in figure \ref{fig:intersection}).

Hence the property {\it ``P''} is preserved, and we can iterate.

\medskip

\begin{figure}[hbt]
\centerline{
\includegraphics[width=60mm]{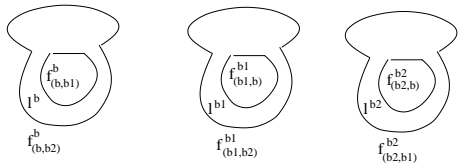}}
\caption{The fourth case.}\label{fig:caz4}
\end{figure}

{\bf Case 4} is depicted in figure \ref{fig:caz4}.

The integral corresponding to $h_{L^b}$ is
\bea
\int dh_{L^b} \delta(h_{L^b}) \delta(h_{L^b}) \delta(h_{L^b} )=
\delta(\id)^2 \; .
\eea

Note that at this point we have exhausted all the lines in a 3D connected component. For every bubble we factored out either by case 2 or by case 3 we obtained a $\delta(\id)$ factor. For the last three bubbles (case 4) we obtained only a $\left(\delta(\id)\right)^2$ factor. We conclude that the amplitude of a ``type 1'' manifold graph is
\bea
    A_{{\cal G}}\,=\,\left(\delta^{\Lambda}(\id)\right)^{|{ B}_{\cal G}|-1}
\eea
\hfill $\square$

We have thus etablished the complete power counting of divergences for the type 1 graphs. Notice that in general, one expects \cite{boulatov,ooguri} the amplitude for the Boulatov model to be given by some divergent factor (in absence of regularization) times a function of the combinatorial structure of the graph and on the specific group chosen (here $SU(2)$) which is a topological invariant of the (pseudo-)manifold it corresponds to. Our result seems to suggest that type 1 graphs are those for which this invariant evaluates to one. For example, in the Turaev-Viro regularization, based on $SU(2)_q$ with $q$ a root of unity, this invariant would be one if and only if the graphs provide a cellular decomposition of a 3-sphere; if this would hold in our setting as well, it would imply that type 1 graphs correspond to manifolds of trivial topology (i-e spheres), thus lending further support to our second conjecture that they should dominate in some appropriate scaling limit, as this is exactly what happens in the similar matrix models. However, the relation between the (divergent) Boulatov/Ponzano-Regge model an the Turaev-Viro model is non-trivial, and such conclusion cannot be drawn so easily in our case.

\section{Conclusions}

In this paper, we have made the first steps in a systematic study of GFT renormalization, focusing on the simple and well-known Boulatov model for 3D quantum gravity. We have thus developed tools that should be later applied to other models, and identified some of the difficulties involved in applying renormalization ideas to GFTs. This 3D model, in fact, has proven to be highly non-trivial, from the point of view of renormalization, due to the complicated combinatorial and topological structure of its Feynman diagrams, given by 3D cellular complexes. Of course, this makes it also a very interesting special type of quantum field theory. 

The divergences of this model come from bubbles, i.e. from the 3D cells of the Feynman diagrams, dual to vertices of the corresponding 3D simplicial complexes. As a first result, we have defined a precise algorithm for constructing the 2D triangulations that characterize the boundary of the 3D bubbles of an arbitrary 3D GFT Feynman diagram.

Introducing an explicit cut-off in the spectrum of the propagator, we have then shown that the only family of graph for which a full contraction procedure exists, and that can thus be considered local from the point of view of renormalization, are the ``type 1'' manifolds defined in section \ref{sec:bubbles}. This means that, in order to be able to renormalize this model, we should be able to find a regime in which all divergent diagrams are ``type 1''. 

For these type 1 diagrams, we have finally proven a power counting theorem. As it is the case with matrix models, this power counting is not uniform in the number of internal vertices. We then conjecture that the regime in which these diagrams are the dominant ones, and where a proper GFT renormalization can be defined, is to be obtained by means of an appropriate scaling limit. We have also exhibited a counterexample showing that this result can not be extended to arbitrary (pseudo-)manifolds.

These first steps in understanding GFT renormalization serve to clarify several key aspects of the physics of GFT models and of non-perturbative quantum gravity in general. Most important, as discussed in the introduction, the origin of the continuum and manifold-like appearance of quantum spacetime at low energies, and of its topology.

\section*{Acknowledgements}
We thank J. Magnen, V. Rivasseau and M. Smerlak for discussions.
D. O. gratefully acknowledges financial support from the Alexander Von Humboldt Foundation through a Sofja Kovalevskaja Prize.

Research at Perimeter Institute is supported by the Government of Canada through Industry 
Canada and by the Province of Ontario through the Ministy of Research and Innovation.

\end{document}